\documentclass[a4paper,11pt]{article}
\usepackage{jcappub}
\usepackage{amsmath}
\usepackage{graphicx}
\usepackage{multirow}

\usepackage{xcolor}

\makeatletter
\gdef\@fpheader{}
\makeatother

\newlength{\fullw}
\newlength{\halfw}
\newlength{\threefigw}
\newlength{\twofigw}
\newlength{\onefigw}
\newlength{\roww}
\newcommand{\order}[1]{\mathcal{O}\!\left(#1\right)}

\newcommand{\hypergauss}[4]{\phantom{}_{_2}\mathrm{F}\!_{_1}\!\left(#1,#2;#3;#4\right)}
\newcommand{\heaviside}[1]{\mathrm{\Theta}\!\left( #1 \right)}
\newcommand{\heavisideb}[1]{\mathrm{\Theta}\!\left[ #1 \right]}
\newcommand{\lpf}[1]{\mathcal{S}\negthinspace\left( #1 \right)}
\newcommand{\lpfb}[1]{\mathcal{S}\negthickspace\left[ #1 \right]}
\newcommand{\func}[2]{f_\mathrm{#1}\!\left(#2\right)}
\newcommand{\funcs}[1]{\func{}{#1}}
\newcommand{\funcc}[1]{\func{c}{#1}}
\newcommand{\funci}[1]{\func{\infty}{#1}}
\newcommand{\gunc}[2]{g_\mathrm{#1}\!\left(#2\right)}
\newcommand{\guncs}[1]{\gunc{}{#1}}
\newcommand{\guncrad}[1]{\gunc{\urad}{#1}}
\newcommand{\guncmat}[1]{\gunc{\umat}{#1}}

\newcommand{\negleft}{\negthinspace\left}
\newcommand{\dirac}[1]{\delta\negleft(#1\right)}

\newcommand{\ud}{\mathrm{d}}
\newcommand{\uc}{\mathrm{c}}
\newcommand{\uh}{\mathrm{h}}

\newcommand{\ugw}{\mathrm{gw}}

\newcommand{\umat}{\mathrm{mat}}
\newcommand{\urad}{\mathrm{rad}}
\newcommand{\uini}{\mathrm{ini}}

\newcommand{\ucrit}{\mathrm{crit}}
\newcommand{\ulss}{\mathrm{lss}}

\newcommand{\uir}{\mathrm{ir}}

\newcommand{\uM}{\mathrm{M}}
\newcommand{\uR}{\mathrm{R}}

\newcommand{\calF}{\mathcal{F}}

\newcommand{\calN}{\mathcal{N}}
\newcommand{\calR}{\mathcal{R}}
\newcommand{\calP}{\mathcal{P}}
\newcommand{\calFini}{\calF_{\uini}}
\newcommand{\Fini}{\calFini}

\newcommand{\calNini}{\calN_\uini}
\newcommand{\calFi}{\calF_{\infty}}

\newcommand{\horizon}{d_\uh}

\newcommand{\chic}{\chi_\uc}
\newcommand{\chimat}{\chi_{_\uM}}
\newcommand{\chirad}{\chi_{_\uR}}
\newcommand{\chicrit}{\chi_{_\ucrit}}
\newcommand{\chiir}{\chi_{\uir}}
\newcommand{\chii}{\chi_{_\infty}}

\newcommand{\gammad}{\gamma_\ud}
\newcommand{\gammac}{\gamma_\uc}
\newcommand{\gammax}{\gamma_\tau}
\newcommand{\gammap}{\gamma_{+}}
\newcommand{\gammai}{\gamma_\infty}
\newcommand{\gammair}{\gamma_\uir}

\newcommand{\muc}{\mu_\uc}
\newcommand{\mui}{\mu_\infty}

\newcommand{\ellc}{\ell_\uc}
\newcommand{\elld}{\ell_\ud}
\newcommand{\ellp}{\ell_{+}}

\newcommand{\ci}{c_\infty}
\newcommand{\cc}{c_\uc}

\newcommand{\zlss}{z_\ulss}

\newcommand{\zini}{z_\uini}

\newcommand{\tini}{t_\uini}
\newcommand{\tx}{t_\tau}
\newcommand{\tp}{t_{+}}
\newcommand{\tc}{t_{\uc}}
\newcommand{\aini}{a_\uini}

\newcommand{\GU}{GU}
\newcommand{\GUU}{GU^2}
\newcommand{\Pgw}{P_\ugw}

\begin{document}

\dedicated{KCL-PH-TH/2019-19}

\title{Cosmic string loop production functions}

\author[a]{Pierre Auclair,}
\author[b]{Christophe Ringeval,}
\author[c]{Mairi Sakellariadou}
\author[a]{and Dani\`ele Steer}

\affiliation[a]{Laboratoire Astroparticule et Cosmologie, Universit\'e
  Denis Diderot Paris 7, 10 rue Alice Domon et Léonie Duquet, 75013
  Paris, France}

\affiliation[b]{Cosmology, Universe and Relativity at Louvain,
  Institute of Mathematics and Physics, Louvain University, 2 Chemin
  du Cyclotron, 1348 Louvain-la-Neuve, Belgium}

\affiliation[c]{Department of Physics, King's College, University of
London, Strand, London WC2R 2LS, United Kingdom}

\emailAdd{auclair@apc.in2p3.fr}
\emailAdd{christophe.ringeval@uclouvain.be}
\emailAdd{mairi.sakellariadou@kcl.ac.uk}
\emailAdd{steer@apc.in2p3.fr}

\date{today}

\abstract{Numerical simulations of Nambu-Goto cosmic strings in an
  expanding universe show that the loop distribution relaxes to an
  universal configuration, the so-called scaling regime, which is of
  power law shape on large scales. Precise estimations of the power
  law exponent are, however, still matter of debate while numerical
  simulations do not incorporate all the radiation and backreaction
  effects expected to affect the network dynamics at small scales. By
using a Boltzmann approach, we show that the steepness of the loop
production function with respect to loops size is associated with
drastic changes in the cosmological loop distribution. For a scale
factor varying as $a(t) \propto t^\nu$, we find that sub-critical loop
production functions, having a Polchinski-Rocha exponent $\chi < (3
\nu - 1)/2$, yield scaling loop distributions which are mostly
insensitive to infra-red (IR) and ultra-violet (UV) assumptions about
the cosmic string network. For those, cosmological predictions are
expected to be relatively robust, in accordance with previous
results. On the contrary, critical and super-critical loop production
functions, having $\chi \ge (3\nu-1)/2$, are shown to be IR-physics
dependent and this generically prevents the loop distribution to relax
towards scaling. In the latter situation, we discuss the additional
regularisations needed for convergence and show that, although a
scaling regime can still be reached, the shape of the cosmological
loop distribution is modified compared to the naive
expectation. Finally, we discuss the implications of our findings.}

\keywords{Cosmic Strings, Loop Production Function, Gravitational Waves}

\maketitle

\section{Introduction}
\label{sec:intro}

The advent of gravitational wave astronomy provides an unprecedented
opportunity to search for topological defects, and in particular
cosmic strings~\cite{Kirzhnits:1972, Kibble:1976, Witten:1985fp,
  Dvali:1998pa}.  In an expanding and decelerating universe, a cosmic
string network relaxes towards an attractor configuration exhibiting
universal properties --- known as a scaling solution --- and it
subsequently remains self-similar with the Hubble
radius~\cite{Hindmarsh:1994re,Vilenkin:2000jqa,Durrer:2002,
  Polchinski:2004ia, Davis:2008dj, Copeland:2009ga,
  Sakellariadou:2009ev, Ringeval:2010ca,Vachaspati:2015cma}. Hence if
cosmic strings were formed in phase transitions early in the history
of the universe, scaling implies that they should be present all over
the sky with a surface density growing with redshift $z$. Strings
induce anisotropies in the Cosmic Microwave Background (CMB) and they
have been searched for in the Planck data~\cite{Ade:2013xla,
  Lazanu:2014eya, Lizarraga:2016onn, 2017MNRAS.472.4081M,
  Sadr:2017hfm, 2019MNRAS.483.5179C}. The current CMB constraints give
an upper bound for the string energy per unit length $U$ of $\GU <
\order{10^{-7}}$, where $G$ is the Newton's constant. However, CMB
photons come from the highest observable redshift set by their last
scattering surface, namely $\zlss \simeq 1088$. For gravitons, $z$ is
only bounded by our understanding of the Friedmann-Lema\^itre model,
or more probably by the redshift at which cosmic inflation ended. For
this reason, the stochastic gravitational wave background (SGWB) is an
observable particularly sensitive to cosmic strings and could provide
the opportunity for a first detection.

Current constraints on $GU$ from the SGWB are already much stronger
than those from the CMB, of order $GU <
\order{10^{-11}}$~\cite{Ringeval:2017eww,
  Blanco-Pillado:2017rnf,Abbott:2017mem} (the actual value depends on
some yet unknown microphysical parameters). However, as opposed to the
CMB constraints, bounds from GW crucially depend on the loop
distribution. Indeed, through their production by the string network,
oscillating closed cosmic string loops constitute the main source of
the SGWB. Although loop production is observed and measured in
Nambu-Goto cosmic string simulations~\cite{Ringeval:2005kr,
  Vanchurin:2005pa, Martins:2005es}, it is still a matter of debate if
it plays the same role in a field theoretical model~\cite{Vincent:1998,
  Moore:2002, Hindmarsh:2008dw, Hindmarsh:2017qff}. Clearly the
detailed shape of the scaling loop distribution function is important
to determine the properties of the SGWB at different
frequencies. Nambu-Goto simulations from two independent groups have
shown that, on \emph{large scales} (see discussion below), where these
simulations can be trusted, it is a power-law, namely
\begin{equation}
t^4 \calF(\gamma,t) \propto \gamma^{p}.
\label{eq:gammabig}
\end{equation}
Here we have defined
\begin{equation}
  \gamma(\ell,t) \equiv \dfrac{\ell}{t}\,, \qquad \calF(\gamma,t)
  \equiv \dfrac{\ud n}{\ud \ell}\,,
  \label{Fdef}
\end{equation}
where $n(\ell,t)$ is the number density distribution of loops of size
$\ell$ at cosmic time $t$, and the time-independence of the
combination $t^4\calF$ is precisely the scaling regime.  The
simulations of Ref.~\cite{Ringeval:2005kr} give
\begin{equation}
  \left. p = -2.60^{-0.21}_{+0.15}\right|_\urad \,, \qquad \left. p
  = -2.41^{-0.08}_{+0.07} \right|_\umat.
\label{eq:p}
\end{equation}
Analysis of the simulations of Refs.~\cite{Blanco-Pillado:2013qja,
  Blanco-Pillado:2017oxo} favours slightly different values, namely
$p=-5/2$ in the radiation and $p=-2$ in the matter era. It
is, however, important to stress that the approach taken in the
numerical simulations of Refs.~\cite{Blanco-Pillado:2013qja,
  Blanco-Pillado:2017oxo} is quite different to that of
Ref.~\cite{Ringeval:2005kr}. In the latter reference, the shape of the
scaling loop distribution $t^4 \calF(\gamma)$ is estimated from
simulations whereas in the former references this is the shape of the
scaling loop production function which is inferred from numerical
results.

Let us also notice that, due to the huge disparity of scales in the
problem (ranging from, for instance, the distance between kinks formed
by string intercommutations, to the horizon size), numerical
simulations of cosmic string networks cannot incorporate all physical
effects. In Nambu-Goto simulations, in particular, effects from GW
emission and backreaction onto the string dynamics are
ignored\footnote{See, however, Ref.~\cite{Quashnock:1990wv} and more
  recently Ref.~\cite{Helfer:2018qgv} for an isolated loop.}.  This is
why Eq.~\eqref{eq:gammabig} can only be trusted for loops large enough
that these effects remain negligible. GW emission means that loops
loose energy and hence become smaller, with an average emitted GW
power $\Pgw = \Gamma \GUU$ where $\Gamma$ is a numerical constant
estimated to be $\Gamma =\order{50}$ ~\cite{Vachaspati:1984gt,
  PhysRevD.45.1898, Blanco-Pillado:2017oxo}. Hence loops decoupled
from the Hubble flow shrink at an average rate given by
\begin{equation}
 \gammad \equiv \Gamma \GU.
\label{eq:gammad}
\end{equation}
One therefore expects Eq.~(\ref{eq:gammabig}) to hold for loops of
length $\ell \gtrsim \elld = \gammad t$ (numeric-wise, this is a quite
small number already for $\GU < 10^{-7}$). Emitted GWs will also
backreact onto the string thereby affecting its dynamics. The
consequences of this process for the network and the loops are still
unknown and being studied~\cite{Wachter:2016rwc}. However, one expects
that loop production should be cut-off below some length scale $\ellc
\equiv \gammac t$, with presumably $\gammac \le \gammad$, which we
discuss below.

As was realised very early on.~\cite{Vilenkin:2000jqa}, in practise, to
include these physical effects one needs to combine results of
simulations with analytical modelling. A powerful framework for this
is to use a Boltzmann approach to estimate the loop distribution on
cosmological time and length scales~\cite{Copeland:1998na,
  Rocha:2007ni, Lorenz:2010sm, Vanchurin:2011hm, Peter:2013jj,
  Vanchurin:2013tk, Schubring:2013xza}. At this stage it is remarkable
to notice that \emph{radically different} assumptions about the loop
production function can lead to \emph{similar} powers $p$ on large
scales (where the results should be fitted against
simulations). Indeed, on one hand, motivated by the one-scale model of
cosmic string evolution~\cite{Kibble:1976,Vilenkin:2000jqa}, an often
studied case is one in which~\cite{Caldwell:1991jj, Damour:2000wa,
  DePies:2007bm, Regimbau:2011bm, Binetruy:2012ze, Kuroyanagi:2012wm,
  Aasi:2013vna, Henrot-Versille:2014jua, Sousa:2016ggw}
\begin{equation}
  \calP(\gamma,t) \propto \dirac{\gamma - \alpha},
\label{eq:delta}
\end{equation}
namely all stable loops are formed with size $\ell = \alpha t$ at time
$t$ (for constant $\alpha$).  It is then straightforward to extract
the loop density distribution~\cite{Vilenkin:2000jqa} (see
Section~\ref{sec:deltaf}) and show that in the radiation era $p=-5/2$
while in the matter era $p=-2$. On the other hand, all cosmic string
simulations show that a lot of small-scale structure, namely kinks
generated from string intercommutation, build up on the strings (see
Refs.~\cite{Bennett:1989, Bennett:1990, Allen:1990, Albrecht:1989,
  Sakellariadou:1990nd,Copeland:2009dk,Austin:1993rg} for a discussion
of small-scale structure on strings). As a result, one expects loops
to be formed on a wide range of scales at any given time. The most
recent analytical work along these lines is by Polchinski-Rocha and
collaborators~\cite{Polchinski:2006ee, Dubath:2007mf, Rocha:2007ni},
who proposed a model of loop production from long strings. It is given
by
\begin{equation}
  t^5 \calP(\gamma>\gammac,t) \propto \gamma^{2\chi-3},
\label{eq:PRchi}
\end{equation}
where the parameter $\chi$ will be referred to as the Polchinski-Rocha
(PR) exponent\footnote{The PR exponent is related to the two-point
  correlation function of tangent vectors along cosmic strings.}. This
is clearly very different from a Dirac distribution as a loop
production function. In Ref.~\cite{Lorenz:2010sm}, the authors have
included backreaction effects to the PR model and extended
Eq.~\eqref{eq:PRchi} to the domains $\gamma < \gammac$, but, motivated
by the numerical results of Ref.~\cite{Ringeval:2005kr}, have
considered only the cases $\chi < \chicrit$ where
\begin{equation}
  \chicrit=\dfrac{3 \nu-1}{2}\,.
  \label{eq:chicrit}
\end{equation}
Here, we have assumed that the scale factor behaves as $a \propto
t^\nu$ so that $\chicrit = 0.25$ and $\chicrit=0.5$ for the radiation
and matter era, respectively. Under the condition $\chi < \chicrit$,
Refs.~\cite{Rocha:2007ni, Lorenz:2010sm} have shown that the loop
distribution behaves as a power law on large scales, with the power
$p$ in Eq.~\eqref{eq:gammabig} given by
\begin{equation}
p=2\chi-3.
\label{eq:pchi}
\end{equation}
From Eqs.~\eqref{eq:p} and \eqref{eq:pchi}, the Nambu-Goto simulations
of Ref.~\cite{Ringeval:2005kr} therefore give
\begin{equation}
\chirad = 0.200^{+0.07}_{-0.10}\,, \qquad \chimat = 0.295^{+0.03}_{-0.04},
\label{eq:chirsb}
\end{equation}
for the radiation and matter era, respectively. We also note that
$\chi$ has been estimated from the two-point correlators of tangent
vectors along the long strings using an average over multiple Abelian
Higgs simulations in Ref.~\cite{Hindmarsh:2008dw} where it was found
that $\chirad=0.22$ and $\chimat=0.35$. At this stage it is intriguing
to notice that the powers $p=-5/2$ in the radiation era, and $p=-2$ in
the matter era, correspond precisely to $\chi=\chicrit$ where the
analysis of Ref.~\cite{Lorenz:2010sm} breaks down. One of the aims of
this paper is precisely to extend the analysis of
Ref.~\cite{Lorenz:2010sm} to the ``critical case'' $\chi = \chicrit$
and to the ``super-critical case'' $\chi > \chicrit$.

Before doing so, however, it is important to comment that while the
two loop production functions of Eqs.~\eqref{eq:delta} and
\eqref{eq:PRchi} lead to similar loop distributions on large scales,
they lead to very important differences for small loops, namely for
$\gamma < \gammad$. Until recently, these differences on small scales
were of no great concern for observable predictions. For instance,
predictions for the CMB power spectrum and induced non-Gaussianities are
essentially blind to cosmic string loops\footnote{The tri-spectrum
  depends however on $\chi$ due to its sensitivity to tangent vector
  correlators ~\cite{Hindmarsh:2009qk, Hindmarsh:2009es}.} (see
Ref.~\cite{Ringeval:2010ca} for a review). However, the situation is
not the same for gravitational waves. The Pochinski-Rocha (PR) loop
production function induces a larger population of small loops. Small
loops oscillate faster, and being more numerous, they can potentially
dominate the GW emission within some frequency range.

In this paper, we show that the value of $\chi=\chicrit$ is a
separatrix between two different behaviours. For values $\chi <
\chicrit$, we recover the results presented in
Refs.~\cite{Lorenz:2010sm, Ringeval:2017eww} and confirm the weak
dependence of the scaling loop distribution on the details of the
backreaction cut-off at small scales. We will refer to this property
as being ultra-violet (UV) insensitive. We also show that the
predicted loop number density is not affected by assumptions made for
the distribution of the largest loops, and this property will be
referred to as infrared (IR) insensitive. On the contrary, values of
$\chi \ge \chicrit$, including the equality, exhibit a very strong
sensitivity to the IR. In fact, under the simplest assumptions, we
show that the loop distribution cannot even reach a scaling regime and
diverges in time. Scaling solutions can still be reached provided
additional assumptions are made to regularise the IR behaviour, the
validity of which still remains to be assessed in the cosmological
context. For all these possible regularised scaling solutions, we show
that the loop distribution shape is modified compared to the naive
expectation.

The paper is organised as follows. In the next section, we recap the
hypothesis and solutions of the Boltzmann equation presented in
Ref.~\cite{Lorenz:2010sm}. We then show in section~\ref{sec:supercrit}
that the solutions can be readily extended to the super-critical cases
$\chi > \chicrit$ and that the loop distribution never reaches scaling
in that case. In section~\ref{sec:crit}, we solve the Boltzmann
equation for the critical value $\chi=\chicrit$ and show again that
the loop distribution diverges with time. In
section~\ref{sec:regularization}, we discuss the extra-assumptions
needed in the IR to produce a scaling loop distribution with $\chi \ge
\chicrit$. For those, we derive the new scaling loop distributions and
critically compare the results in all three cases, sub-critical,
critical and super-critical. We finally conclude by briefly discussing
the implications of our findings.

\section{Cosmic string loop evolution}
\label{sec:evolution}

\subsection{Boltzmann equation and loop production function}

The number density $n(\ell,t)$ of cosmic string loops of size $\ell$
at cosmic time $t$ is assumed to follow a conservation equation
\begin{equation}
\label{eq:dpart}
\dfrac{\ud}{\ud t} \left(a^3 \dfrac{\ud n}{\ud \ell} \right)
= a^3 \calP(\ell,t),
\end{equation}
where $\calP(\ell,t)$ is a loop production function (LPF) giving the
number density distribution of loops of size $\ell$ produced per unit
of time at $t$ and $a(t)$ is the scale factor\footnote{This
  equation can be generalised to include collision terms describing
  loop fragmentation as well as loop collisions, see
  Ref.~\cite{Copeland:1998na}.}.  For an individual loop,
gravitational wave emission induces energy loss through
\begin{equation}
\dfrac{\ud \ell}{\ud t} = -\gammad.
\label{eq:shrink}
\end{equation}
Combining Eqs.~(\ref{eq:dpart}) and (\ref{eq:shrink}), and working in
terms of the variables $(\gamma,t)$ and $\calF\equiv \ud n/\ud \ell$
given in Eq.~(\ref{Fdef}), one obtains the two-dimensional Boltzmann
equation
\begin{equation}
t \dfrac{\partial(a^3 \calF)}{\partial t} - \left(\gamma + \gammad
\right) \dfrac{\partial(a^3 \calF)}{\partial \gamma} = a^3 t
\calP(\gamma,t).
\label{eq:evolgam}
\end{equation}
Its general solution can be obtained by changing variables to $(t,v)$
where $v = t(\gamma + \gammad)$. Then Eq.~(\ref{eq:evolgam}) becomes
\begin{equation}
\label{eq:theone}
    \left. \frac{\partial[a^3\calF(t,v)]}{\partial t}\right|_v = a^3
    \calP(t,v).
\end{equation}
Assuming the infinite (super-horizon) string network is in
scaling, the $t$-dependence of the LPF is of the form
\begin{equation}
t^5 \calP(\gamma,t) = \lpf{\gamma} = \lpf{\frac{v}{t}-\gammad},
\end{equation}
and it is straightforward to integrate Eq.(\ref{eq:theone}) from some
initial time $\tini$ and find its general solution. In terms of the
variables $(\gamma,t)$ it reads
\begin{equation}
    \label{eq:generalsol}
    \calF(\gamma,t) - \Fini(\gamma,t) = \int_{\tini}^{t}
    \left[\frac{a(t')}{a(t)}\right]^3
    \lpfb{\dfrac{(\gamma+\gammad)t}{t'} - \gammad}
      \dfrac{\ud t'}{t'^5},
\end{equation}
where
\begin{equation}
  \calFini(\gamma,t) = \left[\frac{a(\tini)}{a(t)}\right]^3
  \calNini\left[{(\gamma+\gammad)t}-\gammad \tini \right],
\label{eq:Fini}
\end{equation}
with $\calNini(\ell)$ the initial loop distribution at
$t=\tini$. Notice that the time dependence appears because
$\calFini(\gamma,t)$ is evaluated at $t'=\tini$ and physically encodes
the fact that, at time $t$, a loop of length $\gamma t$ corresponds to
an initial loop of size $\ell=\gamma t + \gammad(t-\tini)$. Hence,
once the loop production function $\lpf{\gamma}$ is specified over its
entire domain of definition, the loop distribution is uniquely given
by Eq.~(\ref{eq:generalsol}). As mentioned in the Introduction,
physically very different LPF can give similar loop distributions for
large loops. We now discuss the LPF.

\subsection{Dirac distribution for the loop production function}
\label{sec:deltaf}

In order to compare with results in the literature, let us solve
explicitly the Boltzmann equation with a delta function LPF, motivated
by the one-scale model, given in Eq.~(\ref{eq:delta}), namely
$t^5\calP(\gamma,t) = c\delta(\gamma-\alpha)$. From
Eq.~(\ref{eq:generalsol}),
 \begin{equation}
    t^4 \calF(\gamma < \alpha,t) - t^4 \Fini(\gamma,t) =
    c\left[\dfrac{a\left(t\dfrac{\gamma+\gammad}{\alpha+\gammad}\right)}{a(t)}\right]^3
    \dfrac{(\alpha+\gammad)^3} {(\gamma+\gammad)^{4}}
    \heavisideb{\gamma + \gammad - \dfrac{\tini}{t}(\alpha+\gammad)}.
    \label{eq:deltaf}
\end{equation}
The left-hand side of Eq.~\eqref{eq:deltaf} contains $\Fini$, which is
determined from the initial loop distribution $\calNini$ through
Eq.~(\ref{eq:Fini}).  This term is usually a transient for initial
loop distribution converging fast enough to zero at large
$\ell$. However, if (as in numerical simulations) $\calNini$ is
assumed to be the Vachaspati-Vilenkin (VV)
distribution~\cite{Vachaspati:1984dz} one has $\tini^4 \calNini(\ell)
\propto (\tini/\ell)^{5/2}$ and because the argument of $\calNini$ in
Eq.~\eqref{eq:Fini} grows with $t$ we see that, in the particular case
of the radiation era ($\nu=1/2$), the whole term becomes
time-independent and ``scales''. In a realistic situation, the VV
distribution is valid up to some size, typically the initial horizon
size $\ell < \horizon(\tini)$, where $\horizon(t)=t/(1-\nu)$ with
$\nu=1/2$ or $2/3$ in the radiation or matter era, respectively. Above
$\horizon(\tini)$, loops are of super-horizon length and should
actually be considered as long (dubbed ``infinite'') strings from a
dynamical point of view. Once the argument of $\calNini$ (through
$\Fini$) in Eq.~\eqref{eq:deltaf} becomes larger than this cut-off,
the corresponding term in the left-hand side of Eq.~\eqref{eq:deltaf}
disappears.

Neglecting therefore the effects from initial distribution
$\Fini(\gamma,t)$, we find the loop distribution in the radiation era:
  \begin{equation}
      t^4 \calF(\gamma,t) = c\,\dfrac{(\alpha+\gammad)^{3/2}}
  {(\gamma+\gammad)^{5/2}} \heaviside{\alpha-\gamma}.
\end{equation}
This expression corresponds to a scaling solution with a $p=-5/2$
power-law for $\gamma\gg \gammad$, as stated in the Introduction. For
loops formed during matter era one has
\begin{equation}
  t^4 \calF(\gamma,t) = c\,\dfrac{(\alpha+\gammad)}
  {(\gamma+\gammad)^{2}} \heaviside{\alpha-\gamma},
\end{equation}
and this corresponds to a scaling solution with a $p=-2$ power-law for
$\gamma\gg \gammad$. Notice that, in both cases, the distributions are
flat for values of $\gamma < \gammad$.

\subsection{Polchinksi-Rocha loop production function}

\begin{figure}
  \begin{center}
    \includegraphics[width=\onefigw]{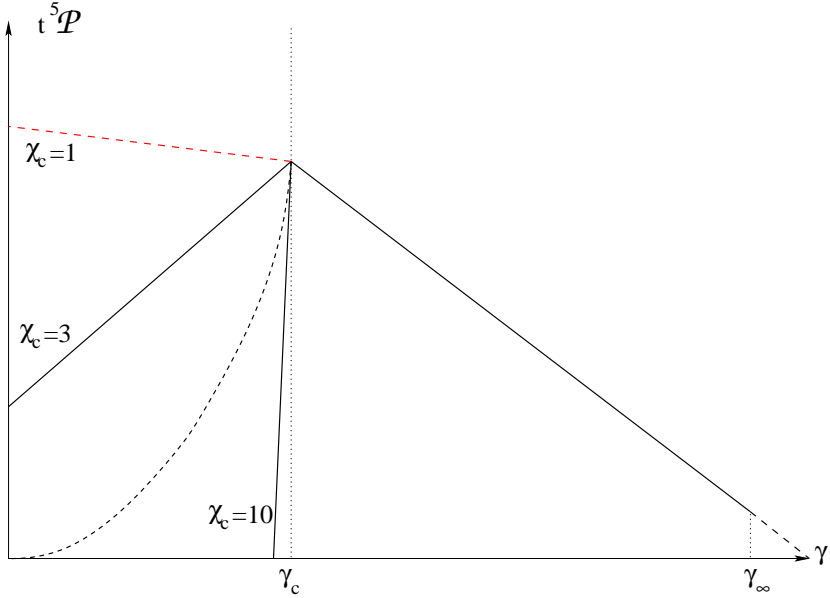}
    \caption{Sketch of possible loop production function shapes under
      the gravitational backreaction length scale
      $\gammac\equiv\ellc/t$ (logarithmic units), namely $\calP(\gamma
      \le \gammac,t) = \cc\, \gamma^{2\chic-3}$ where the constant
      $\cc$ is chosen such that $\calP$ is continuous at
      $\gamma=\gammac$. According to Ref.~\cite{Polchinski:2007rg},
      minimal gravitational backreaction effects correspond to
      $\chic=1$ and we take this value as a motivated lower bound. The
      larger the value of $\chic$, the sharper the cut is.}
    \label{fig:lpsketch}
  \end{center}
\end{figure}

In the remainder of this paper we focus on the PR loop production
function, which exhibits a power-law dependence in $\gamma$. For large
loops, it is given by
\begin{equation}
t^5 \calP(\gamma \ge \gammac,t) = c\,\gamma^{2\chi-3}.
\label{eq:Pbig}
\end{equation}
The ``backreaction scale'' $\gammac$ was calculated in
Ref.~\cite{Polchinski:2007rg} and is given by\footnote{The dependence
  on $GU$ is to be expected given that this scale is fixed by
  gravitational physics.}
\begin{equation}
\gammac \equiv \Upsilon (\GU)^{1+2\chi},
\label{eq:gammac}
\end{equation}
where $\Upsilon =\order{20}$. This suggests that the very small scales
on a string network can potentially be strongly dependent on the value
of $\chi$.  On scales $\gamma < \gammac$, the actual shape of the LPF
is unknown, but, surely, loop production has to be cut-off. A
phenomenologically motivated expression has been proposed in
Ref.~\cite{Lorenz:2010sm}, namely
\begin{equation}
t^5 \calP(\gamma < \gammac,t) = \cc\, \gamma^{2\chic-3},
\label{eq:Psmall}
\end{equation}
with $\chic > 1$. Continuity of the loop production function at
$\gamma=\gammac$ imposes
\begin{equation}
  \cc = c \, \gammac^{2(\chi - \chic)}.
  \label{eq:crels}
\end{equation}
The scaling function $\lpf{\gamma}$ is completely determined by
Eqs.~\eqref{eq:Pbig} and \eqref{eq:Psmall} and reads
\begin{equation}
\lpf{\gamma} = c\, \gamma^{2 \chi -3} \heaviside{\gamma-\gammac} + \cc
\, \gamma^{2 \chic - 3} \heaviside{\gammac - \gamma}.
\label{eq:lpf}
\end{equation}

Before giving explicit solutions of the Boltzmann equation for the PR
based LPF, let us remark that the original PR model applies to loops
produced by long (dubbed ``infinite'') strings, whereas in numerical
simulations loops are also created from other loops and can
potentially reconnect.  Hence, the fit to numerical simulations can be
viewed as a renormalisation procedure that allows us to extend the
properties of loops chopped off from long strings to those produced by
other loops. In particular, the fit completely fixes the
  normalisation constant $c$ in the loop distribution. Unless
  specified otherwise, we have used the values reported in
  Ref.~\cite{Ringeval:2005kr}. Simulations show that the largest loops
  created in a cosmological network are as large as the largest
  correlation length scale, which is a fraction of the Hubble
  radius. This typical correlation length allows us to define
\begin{equation}
  \gammai = \left(\dfrac{U}{\rho_\infty t^2} \right)^{1/2},
\end{equation}
where $\rho_\infty$ is the energy density of super-horizon sized
(infinite) strings in scaling~\cite{Bennett:1989, Bennett:1990,
  Allen:1990, Ringeval:2005kr}. One gets $\gammai \simeq 0.32$ in the
radiation era and $\gammai \simeq 0.56$ in the matter era. The PR
model with values of $c$ consistent with those of simulations predicts
a fractional number of loops having $\gamma \ge \gammai$. However, and
as sketched in Fig.~\ref{fig:lpsketch}, the IR behaviour of
$\calP(\gamma,t)$ (at large $\gamma$) could a priori be different than
for $\gamma < \gammai$ and we will explore this possibility in
section~\ref{sec:regularization}.

\subsection{Non-critical loop production function}
\label{sec:munonzero}

In this section, we present the solution of the Boltzmann equation
obtained for the non-critical cases, i.e., $\chi \neq \chicrit$. As
shown in Ref.~\cite{Lorenz:2010sm}, substituting Eq.~(\ref{eq:lpf})
into Eq.~(\ref{eq:generalsol}) gives the unique solution. In the domain
$\gamma \ge \gammac$ it reads
\begin{equation}
  \label{eq:solgd}
  t^4 \calF(\gamma \ge \gammac,t) =   t^4\Fini(\gamma,t) +
  \dfrac{c}{\mu}(\gamma + \gammad)^{2\chi-3} \left[
    \funcs{\dfrac{\gammad}{\gamma + \gammad}}  -
    \left(\dfrac{t}{\tini}
    \right)^{-\mu} \funcs{\dfrac{\gammad}{\gamma + \gammad}
      \dfrac{\tini}{t}} \right],
\end{equation}
and, in the domain $\gamma < \gammac$,
\begin{equation}
  \begin{aligned}
    t^4\calF(\gamma < \gammac,t) &= t^4\Fini(\gamma,t) +
    \dfrac{c}{\mu} (\gamma + \gammad)^{3 \nu - 4}
    \left(\gammac+\gammad \right)^{-\mu}
    \funcs{\dfrac{\gammad}{\gammac+\gammad}} \\ & - \dfrac{c}{\mu}
    (\gamma + \gammad)^{2\chi - 3} \left(\dfrac{t}{\tini}
    \right)^{-\mu} \funcs{\dfrac{\gammad}{\gamma + \gammad}
      \dfrac{\tini}{t}} \\ & + \dfrac{\cc}{\muc} (\gamma + \gammad)^{2
      \chic - 3} \left[\funcc{\dfrac{\gammad}{\gamma + \gammad}} -
      \left(\dfrac{\gamma+\gammad}{\gammac+\gammad}\right)^{\muc}
      \funcc{\dfrac{\gammad}{\gammac+\gammad}} \right].
    \label{eq:solgc}
  \end{aligned}
\end{equation}
In these equations, we have defined
\begin{equation}
\funcs{x} \equiv \hypergauss{3-2\chi}{\mu}{\mu+1}{x}, \qquad \funcc{x}
\equiv \hypergauss{3-2\chic}{\muc}{\muc+1}{x}.
\end{equation}
with $\hypergauss{a}{b}{c}{x}$ being the Gauss hypergeometric
function, and
\begin{equation}
\mu \equiv 3 \nu - 2 \chi -1, \qquad \muc \equiv 3 \nu - 2 \chic -1.
\label{nudef}
\end{equation}
The above solution is valid provided one waits long enough for some
transient domains to disappear\footnote{In the matter era, the
  hypergeometric function simplifies to a polynomial expression, see
  Eq.~(55) in Ref.~\cite{Lorenz:2010sm}.}. For completeness, the full
solution including the transients is presented in the
appendix~\ref{app:fullsols}. Let us stress that these equations become
singular for $\mu = 0$, which corresponds to $\chi = \chicrit$, and
that case must be treated separately, see section~\ref{sec:crit}.

The behaviour of the solution given by Eqs.~\eqref{eq:solgc} and
\eqref{eq:solgd} depends on whether $\chi < \chicrit$, which we refer
to as the sub-critical case, or whether $\chi > \chicrit$, the
super-critical one.

\subsubsection{Sub-critical loop production function}
\label{sec:subcrit}

As discussed in section~\ref{sec:deltaf}, the first term in the
right-hand side of Eq.~\eqref{eq:solgd}, which is determined from the
initial loop distribution, vanishes if one waits long enough. For all
positive values of $\mu$, namely $\chi < \chicrit$, the last term in
Eq.~\eqref{eq:solgd} is also a transient that
asymptotically vanishes for $t \gg \tini$. At vanishing argument, the
hypergeometric function converges to unity and the time dependence of
this term indeed scales as $(t/\tini)^{-\mu}$.

Hence the Boltzmann equation for $\mu > 0$ predicts a scaling loop
distribution for $\gamma \ge \gammac$ given by
\begin{equation}
t^4 \calF(\gamma \ge \gammac,t) = \dfrac{c}{\mu} \left(\gamma +
\gammad \right)^{2\chi-3} \funcs{\dfrac{\gammad}{\gamma + \gammad}}.
\label{eq:scalsuperd}
\end{equation}
For $\gamma \gg \gammad$ the hypergeometric function tends to 1, and
we recover the power-law distribution given in Eq.~(\ref{eq:pchi}); it
matches numerical simulations where gravitational effects are absent:
\begin{equation}
t^4 \calF(\gamma \gg \gammad,t) \simeq \dfrac{c}{\mu} \, \gamma^{2\chi-3}.
\label{eq:approxsuperd}
\end{equation}
Furthermore we can now predict the effects associated with
\emph{gravitational wave emission}. Taking the limit $\gamma \ll
\gammad$ (but still $\gamma > \gammac$), one gets\footnote{To derive
  this expression, we have expanded the hypergeometric function around
  unity~\cite{Gradshteyn:1965aa}
\begin{equation}
\funcs{x} \underset{1}{\sim} \dfrac{\Gamma(3 \nu - 2\chi)
  \Gamma(2\chi-2)}{\Gamma(3\nu-3)} x^{-\mu} + \dfrac{\mu}{2-2\chi}
\left(1-x\right)^{2\chi-2}.
\label{eq:hyperone}
\end{equation}
}
\begin{equation}
t^4 \calF(\gammac <\gamma \ll \gammad,t) \simeq \dfrac{c}{2-2\chi}
\dfrac{\gamma^{2\chi-2}}{\gammad}\,.
\label{eq:scalsubd}
\end{equation}
Notice that since we are in the regime $\chi<\chicrit$ we necessarily
have $\chi < 1$. The only effect of gravitational wave emission onto
the scaling loop distribution is to reduce the power law exponent by
one unit in the domain $\gammac < \gamma \ll
\gammad$~\cite{Rocha:2007ni}.

\begin{figure}
  \begin{center}
    \includegraphics[width=\onefigw]{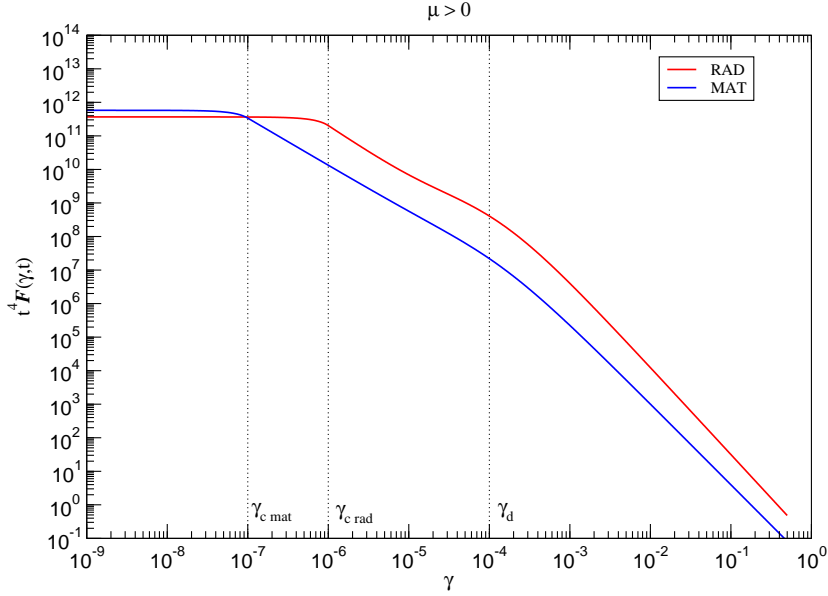}
    \caption{Scaling loop distribution in the radiation and matter
      era for $\mu>0$, which corresponds to $\chi < (3\nu -1)/2$. The
      values for $\gammad$ and $\gammac$ are illustrative only.}
    \label{fig:scalingmupos}
  \end{center}
\end{figure}

To see what are the effects of \emph{gravitational wave backreaction}
on the loop distribution, let us consider Eq.~\eqref{eq:solgc}. As
before, the first and third terms in the right-hand side of
Eq.~\eqref{eq:solgc} are transient and only the second term and the
fourth one survive. They are explicitly time-independent showing that
this part of the loop distribution also reaches scaling. Using the
expansion~\eqref{eq:hyperone}, the matching condition
\eqref{eq:crels}, and taking the limit $\gamma \ll \gammac$ gives
\begin{equation}
\begin{aligned}
  t^4 \calF(\gamma \ll \gammac,t) &= c \left(\dfrac{1}{2-2\chi} +
  \dfrac{1}{2\chic-2} \right) \dfrac{\gammac^{2\chi-2}}{\gammad} +
  \order{\gammad^{2\chi-3}} \\ & \simeq \dfrac{c}{2-2\chi}
  \dfrac{\gammac^{2\chi-2}}{\gammad}\,,
\label{eq:subcritUV}
\end{aligned}
\end{equation}
where in the last step we have taken the limit for $\chic \gg \chi$
and $\gammac \ll \gammad$. This expression makes clear that the
details of the backreaction process, namely the values of $\chic$,
have only a weak effect on the final loop
distribution~\cite{Lorenz:2010sm}. Therefore, in the domain $\gamma <
\gammac$, the scaling loop distribution is flat.

The exact form for the scaling loop distribution is plotted in
Fig.~\ref{fig:scalingmupos} for both the radiation and matter era, see
also Eqs.~\eqref{eq:fullsolgd} to \eqref{eq:fullsolgo}. Notice that
the value of $\gammac$ is $\chi$-dependent, and thus, even at constant
$\GU$, $\gammac$ changes between radiation and matter.

\subsubsection{Super-critical loop production function}

\label{sec:supercrit}

As discussed in the Introduction, we now consider shallower loop
production functions having $\mu < 0$, i.e.~super-critical values of
$\chi > \chicrit$. All solutions derived in
section~\ref{sec:munonzero} are regular in this limit, and we can
straightforwardly use  Eqs.~\eqref{eq:solgd} and \eqref{eq:solgc}.

In the domain $\gamma \ge \gammac$, neglecting the first term in the
right-hand side of Eq.~\eqref{eq:solgd} for the afore-mentioned
reasons, we see that the third term (which was a transient for $\mu >
0$) is now becoming a growing function of time as it scales as
$(t/\tini)^{-\mu}$. Therefore, for $t \gg \tini$, and for all values
of $\gamma \ge \gammac$, the hypergeometric function that multiplies
$(t/\tini)^{-\mu}$ in Eq.~\eqref{eq:solgd} approaches unity and one
gets
\begin{equation}
t^4 \calF(\gamma \ge \gammac,t) \simeq -\dfrac{c}{\mu} \left(\gamma +
\gammad\right)^{2\chi-3} \left[-
  \funcs{\dfrac{\gammad}{\gamma+\gammad}} +
  \left(\dfrac{t}{\tini}\right)^{-\mu} \right],
\label{eq:pathomuneg}
\end{equation}
which is not scaling! Another feature of this solution is that, taking
the limit $\gammac \le \gamma \ll \gammad$, one has
\begin{equation}
t^4 \calF(\gammac \le \gamma \ll \gammad,t) \simeq -\dfrac{c}{\mu}\,
\gammad^{2\chi-3} \left[-\dfrac{\mu}{2-2\chi}
  \left(\dfrac{\gamma}{\gammad}\right)^{2\chi-2} +
  \left(\dfrac{t}{\tini}\right)^{-\mu}\right].
\label{eq:supcritinter}
\end{equation}
The solution only exhibits the $\gamma^{2\chi-2}$ power-law
transiently. As soon as the growing term $(t/\tini)^{-\mu}$ takes
over, the loop distribution becomes flat and incessantly grows with
time. Notice that because $\mu<0$, positiveness of the loop
distribution still implies that $c>0$ because it is now dominated by
the terms $(t/\tini)^{-\mu}$. Equation~\eqref{eq:crels} implies $\cc >
0$ as well.

The solution in the domain $\gamma < \gammac$ presents the same
pathology, namely, the fourth term of  Eq.~\eqref{eq:solgc}, which is a
transient for $\mu>0$, now becomes dominant and one gets for $\gamma
\ll \gammac$
\begin{equation}
t^4 \calF(\gamma \ll \gammac,t) \simeq -\dfrac{c}{\mu}
\gammad^{2\chi-3} \left[-\left(\dfrac{\mu}{2-2\chi} +
  \dfrac{\mu}{2\chic-2} \right)\left(\dfrac{\gammac}{\gammad}
  \right)^{2\chi-2} + \left(\dfrac{t}{\tini}\right)^{-\mu} \right],
\label{eq:supcritsmall}
\end{equation}
which is flat and smoothly connects to the solution
\eqref{eq:supcritinter} at $\gamma=\gammac$.

\begin{figure}
  \begin{center}
    \includegraphics[width=\onefigw]{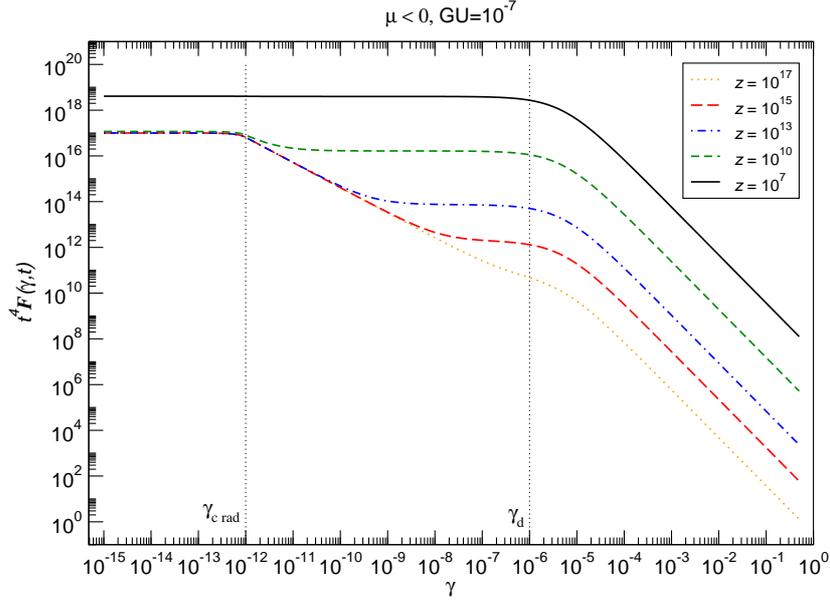}
    \caption{Growing loop distribution generated by a super-critical
      loop production function having $\chi=0.45$ during the radiation
      era. The string tension has been set to $\GU=10^{-7}$ and the
      initial conditions are arbitrarily set at $\zini=10^{18}$ with
      $\calNini(\ell)=0$ and $c=0.14$. At redshift $z=10^{7}$, the
      change of shape associated with gravitational wave backreaction
      becomes washed out by the number loops which diverges with
      time.}
    \label{fig:pathomuneg}
  \end{center}
\end{figure}

In Fig.~\ref{fig:pathomuneg}, we have plotted the exact solutions at
various successive redshifts showing the non-scaling behaviour of the
super-critical cases, $\chi > \chicrit$. The time divergence ends up
washing out the change in slope of the loop distribution between
$\gammac$ and $\gammad$. But scaling is lost and we have an
incessantly growing number density of loops at all scales.

Because Eq.~\eqref{eq:pathomuneg} is actually valid in the regime
probed by numerical simulations, this behaviour not being observed, we
conclude that deeply super-critical loop production functions are
unlikely to be physical. Of course, one cannot exclude the possibility
that $\mu<0$ but very close to zero (hence $\chi$ close to its
critical value $\chic$), since the time-dependence of
Eq.~\eqref{eq:pathomuneg} would remain hardly visible in time-limited
numerical simulations while being relevant on cosmological
time-scales. We now turn to the critical case itself, $\mu=0$.

\subsection{Critical loop production function}
\label{sec:crit}

\begin{figure}
  \begin{center}
    \includegraphics[width=\onefigw]{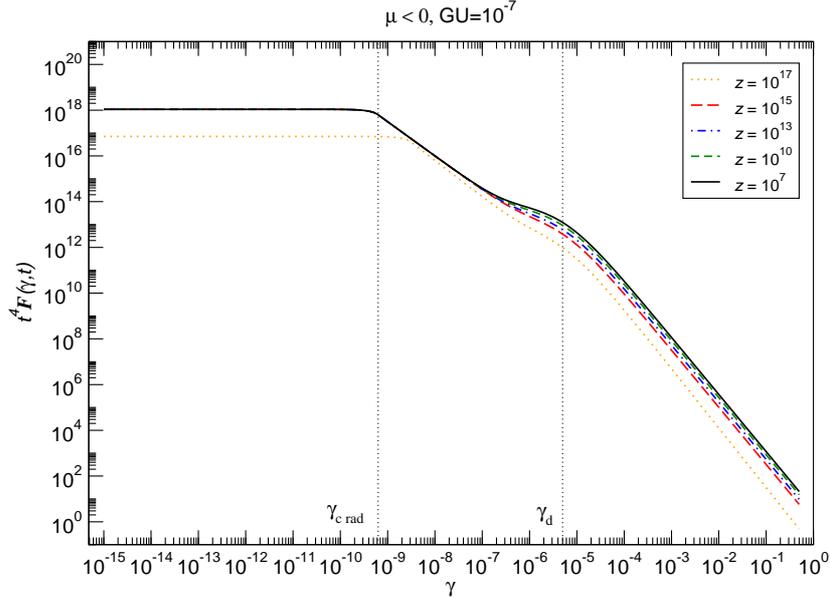}
    \caption{Loop number density distribution at various redshifts for
      a critical loop production function having
      $\chirad=\chicrit=0.25$. The network is assumed to be formed at
      $\zini=10^{18}$ and $c=0.03$. At redshift $z=10^{17}$, the loop
      distribution is not yet fully relaxed from the initial
      conditions. For later redshifts, $z<10^{15}$, the non-scaling
      logarithmic divergence becomes clearly visible for all loops
      larger than the gravitational wave emission scale, $\gamma \ge
      \gammad$. The smaller ones, having $\gamma < \gammad$, remain in
      a transient scaling for most of the cosmological evolution,
      until the non-scaling behaviour takes over (see text).}
    \label{fig:critrad}
  \end{center}
\end{figure}

None of the solutions of section~\ref{sec:munonzero} are valid for
$\mu=0$. Hence we return to the general solution \eqref{eq:generalsol}
where, using Eq.~(\ref{eq:Pbig}) with $\chi=\chicrit$ given in
Eq.~\eqref{eq:chicrit}, one has
\begin{equation}
  \lpf{\gamma}=c \, \gamma^{3\nu-4}\heaviside{\gamma-\gammac}
  + \cc\, \gamma^{2\chic-3}\heaviside{\gammac-\gamma}.
\label{eq:lpfcrit}
\end{equation}
Here we have used the equality $2\chicrit-3 = 3\nu-4$. As before, the
initial condition at $t=\tini$ and continuity of the solution at
$\gamma=\gammac$, which is enforced by Eq.~\eqref{eq:lpfcrit},
completely fix the solution of Eq.~\eqref{eq:generalsol}. We still
find a complete integral (see Ref.~\cite{Gradshteyn:1965aa}) that is
presented, in full, in the appendix~\ref{sec:fullc}. Below, we report
only the parts relevant for our discussion. In the domain $\gamma \ge
\gammac$, one has
\begin{equation}
\label{eq:critgd}
t^4 \calF(\gamma \ge \gammac,t) = t^4\Fini(\gamma,t) + c
(\gamma + \gammad)^{3\nu-4} \left[ \guncs{\dfrac{\gammad}{\gamma + \gammad}}
- \guncs{\dfrac{\gammad}{\gamma + \gammad} \dfrac{\tini}{t}} \right],
\end{equation}
and in the domain $\gamma < \gammac$, the solution reads
\begin{equation}
\begin{aligned}
t^4\calF(\gamma < \gammac,t) &= t^4\Fini(\gamma,t)
 \\ & + \dfrac{\cc}{\muc} (\gamma + \gammad)^{2 \chic - 3}
\left[\funcc{\dfrac{\gammad}{\gamma + \gammad}} -
  \left(\dfrac{\gamma+\gammad}{\gammac+\gammad}\right)^{\muc}
  \funcc{\dfrac{\gammad}{\gammac+\gammad}} \right] \\ & + c
(\gamma + \gammad)^{3\nu-4}
\left[\guncs{\dfrac{\gammad}{\gammac+\gammad}} -
  \guncs{\dfrac{\gammad}{\gamma + \gammad} \dfrac{\tini}{t}} \right].
\label{eq:critgc}
\end{aligned}
\end{equation}
The function $\guncs{x}$ is $\nu$-dependent. In the radiation era, for
$\nu = 1/2$, it reads
\begin{equation}
\guncrad{x} \equiv \ln\left(\dfrac{1-\sqrt{1-x}}{1+\sqrt{1-x}}\right)
+ \dfrac{2}{3} \dfrac{4-3 x}{(1-x)^{3/2}}\,,
\end{equation}
while in the matter era, for $\nu=2/3$,
\begin{equation}
\guncmat{x} \equiv \dfrac{1}{1-x} \ln\left(\dfrac{1-x}{x}\right).
\end{equation}

As before, neglecting the terms associated with $\calNini$, and taking
the limit $t \gg \tini$, Eq.~\eqref{eq:critgd} can be further expanded
for $\gamma \gg \gammad$ as
\begin{equation}
  t^4 \calF(\gamma \gg \gammad,t) \simeq c \, \gamma^{3\nu-4}
  \ln\left(\dfrac{t}{\tini}\right),
\end{equation}
for both the radiation and matter eras. As a result, the critical case
$\chi=\chicrit$ suffers from the same problems as the super-critical
ones: the loop number distribution never reaches a scaling regime. For
$\mu=0$, the power-law exponent is $3 \nu-4 = 2\chicrit-3$ and
smoothly connects to its sub- and super-critical values. Let us notice
however that the time divergence is logarithmic, and therefore, could
very well remain undetected in numerical simulations while being quite
relevant on cosmological time-scales. The limit $\gammac \le \gamma
\ll \gammad$ gives
\begin{equation}
t^4 \calF(\gammac \le \gamma \ll \gammad,t) \simeq c \, \gammad^{3\nu-4}
\left[\dfrac{1}{3-3\nu} \left(\dfrac{\gamma}{\gammad}\right)^{3\nu-3}
  + \ln\left(\dfrac{t}{\tini}\right) \right],
\label{eq:critinter}
\end{equation}
which, up to the logarithmic divergence, is in all points similar to
Eq.~\eqref{eq:supcritinter}. As for the super-critical case, in the
future infinity limit $t/\tini \to \infty$, the dependence in $\gamma$
disappears, the loop distribution becomes flat, grows, and never
reaches scaling. However, because the divergence is only logarithmic in
time, even on cosmological time scales, the first term can remain
dominant. In this situation, we are in presence of a very long
transient scaling in the domain $\gamma \ll \gammad$.

Finally, for the small loops $\gamma \ll \gammac$, and assuming
$\gammac \ll \gammad$, we can expand Eq.~\eqref{eq:critgc} at large
times $t \gg \tini$. We get
\begin{equation}
\begin{aligned}
  t^4 \calF(\gamma \ll \gammac,t) & \simeq \dfrac{\cc}{ 2\chic-2}
  \dfrac{\gammac^{2\chic-2}}{\gammad} + c \gammad^{3\nu-4}
  \left[\dfrac{1}{3 -
      3\nu}\left(\dfrac{\gammac}{\gammad}\right)^{3\nu-3} +
    \ln{\left(\dfrac{t}{\tini} \right)} \right] \\ & = c\,
  \gammad^{3\nu-4} \left[\left(\dfrac{1}{3-3\nu} + \dfrac{1}{2\chic-2}
    \right) \left( \dfrac{\gammac}{\gammad}\right) ^{3\nu-3} +
     \ln{\left(\dfrac{t}{\tini}\right)} \right],
\end{aligned}
\label{eq:critUV}
\end{equation}
where the last step is obtained from Eq.~\eqref{eq:crels}, which
ensures the continuity of the loop production function. Again, this is
in all point similar to the super-critical case of
Eq.~\eqref{eq:supcritsmall} and smoothly connects to the domain
$\gamma \ge \gammac$. The logarithmic divergence will ultimately make
the small loop number density grow, although the presence of the
first term will strongly delay this process and one should expect a
very long transient scaling.

Figure~\ref{fig:critrad} shows the loop number density distribution in
the radiation era as derived from the exact expression,
Eqs.~\eqref{eq:fullcritgd} to \eqref{eq:fullcritgo}, for
$\GU=10^{-7}$, and at various redshifts. Here again, $\calNini=0$ has
been assumed to clearly show the effects coming from the production
function. The network is arbitrarily assumed to be formed at
$\zini=10^{18}$ and relaxation from the initial conditions takes place
down to redshift $z=10^{17}$. For redshifts $z \le 10^{15}$, the domain
$\gamma \ge \gammad$ clearly exhibits the logarithmic divergence. The
loops having $\gamma < \gammad$ remain, however, in the transient
scaling for essentially all the cosmological evolution.

\subsection{Discussion}

Critical and super-critical loop
production functions, having $\chi \ge \chicrit = (3\nu-1)/2$, yield a
non-scaling and growing population of cosmic string loops. This
results from the combination of various non-trivial effects acting
together. For $\chi \ge \chicrit$, the loop production functions are
shallower with respect to loop sizes than the sub-critical
ones. Therefore, they produce, on site, relatively more larger loops
compared to the smaller ones. These larger loops will contribute to
the final population of loops of given size since they incessantly
shrink by gravitational wave emission. Similarly, at all times, loops
of given size disappear by the same effect. The detailed balance of
loops disappearing, being created on site, and being populated by
shrunk larger loops is obviously $\chi$-dependent and the overall
result is precisely given by the solution of the Boltzmann
equation~\eqref{eq:evolgam}. Taking shallower loop production
functions clearly enhances the feeding by larger loops, at all
scales. The critical value $\chicrit$ is the precise power-law
exponent above which such an effect produces a non-stationary
solution. We summarise our results in
Table~\ref{table:dominantnoircutoff}.

\begin{table}
\begin{center}
\renewcommand{\arraystretch}{2}
\begin{tabular}{|l|c|c|c|}
    \hline Type & \(\gamma < \gammac\) & \(\gammac < \gamma <
    \gammad\) & \(\gamma > \gammad\) \\ \hline Sub-critical \(\mu>0\)
    &\(\dfrac{c}{2-2\chi}\gammac^{2\chi-2}\gammad^{-1}\)&\(\dfrac{c}{2-2\chi}\gamma^{2\chi-2}\gammad^{-1}\)&\(\dfrac{c}{\mu}\gamma^{2\chi-3}\)
    \\ Critical \(\mu=0\)&\(c \,
    \gammad^{3\nu-4}\ln\left(\dfrac{t}{\tini}\right)\)&\(c \,
    \gammad^{3\nu-4}\ln\left(\dfrac{t}{\tini}\right)\)&\(c \,
    \gamma^{3\nu-4}\ln\left(\dfrac{t}{\tini}\right)\)
    \\ Super-critical
    \(\mu<0\)&\(-\dfrac{c}{\mu}\gammad^{2\chi-3}\left(\dfrac{t}{\tini}\right)^{-\mu}\)&\(-\dfrac{c}{\mu}
    \gammad^{2\chi-3}
    \left(\dfrac{t}{\tini}\right)^{-\mu}\)&\(-\dfrac{c}{\mu}\gamma^{2\chi-3}
    \left(\dfrac{t}{\tini}\right)^{-\mu}\) \\ \hline
\end{tabular}
\caption{Asymptotic contributions to the loop number density assuming
  no infrared regularisation. At late times, the critical and
  super-critical cases are non-scaling and the loop number density
  diverges. For the critical case, notice however that a transient scaling can take
  place in the domains $\gamma < \gammad$ for most of the cosmological
  evolution (see text).}
\label{table:dominantnoircutoff}
\end{center}
\end{table}

In striking contrast with the sub-critical case, we see that the
critical and super-critical loop production functions induce
non-scaling loop distributions. This is quite dramatic in the
super-critical case as the number density of loops grows, on all
scales, as $(t/\tini)^{-\mu}$, with $\mu < 0$. The situation for the
critical case $\chi=\chicrit$ is, somehow, less catastrophic, as the
divergence is only logarithmic in time. In particular, for most of the
cosmologically relevant situations, we find that the loop number
density remains in a transient scaling regime at small scales, for all
$\gamma \ll \gammad$. The number density of larger loops, having
$\gamma \ge \gammad$, is however logarithmically growing with time and
never scales.

\section{Possible infrared regularisations}
\label{sec:regularization}

In view of the previous discussion, a way to regularise (super-)
critical loop production functions is to change their shape in some
domains. As discussed in the Introduction, the PR model does not
necessarily apply to super-horizon loops, the ones having $\gamma >
\gammai$, and these ones seem to be precisely responsible for the time
divergence. A possible regularisation is therefore making a hard cut
in the IR, namely postulating that the loop production function is
exactly vanishing above some new IR scale, say $\gamma >
\gammai$. Other regulator shapes are considered in
section~\ref{sec:smoothIR}.

We now consider the same PR loop production function as in
section~\ref{sec:evolution} for $\gamma \le \gammai$ but we now
require that $t^4 \calP(\gamma > \gammai,t) = 0$ at all times. As a
result, there is a new domain of solution for Eq.~\eqref{eq:evolgam}
in which one trivially finds
\begin{equation}
  \calF(\gamma \ge \gammai,t) = \Fini(\gamma,t).
\label{eq:reggi}
\end{equation}
The calculations are slightly longer than in
section~\ref{sec:evolution} but do not present new difficulties. They
are detailed in the appendix~\ref{app:fullregs}. The introduction of
a new scale at $\gammai$ introduces various new transient domains in
which the loop distribution $t^4 \calF$ grows for a while before
becoming stationary. Ignoring these domains, the main changes can be
summarised as follows.

The asymptotic solutions are given by those of the previous section
\emph{provided} we make the formal replacement
\begin{equation}
\dfrac{t}{\tini} \longrightarrow \dfrac{\gammai + \gammad}{\gamma + \gammad}\,.
\label{eq:formal}
\end{equation}
This expression makes clear that all terms that were explicitly
depending on $t/\tini$ are regularised to $\gamma$-dependent terms.
As a result, the IR-regularised critical loop distribution reaches
scaling, but it does no longer exhibit the same shape on large
scales. In the following, we explicitly derive the induced distortions
for the critical and super-critical case and discuss the impact of
forcing an unneeded IR-regularisation to the sub-critical loop
production functions.

\subsection{Critical loop production function}
\label{sec:regcritIR}

For critical loop production function $\chi=\chicrit$, after the
disappearance of the transient domains (see
appendix~\ref{app:fullregs}), the loop distribution in the domain
$\gamma \ge \gammac$ (and $\gamma < \gammai$) reads
\begin{equation}
\begin{aligned}
 \label{eq:regcritgd}
t^4 \calF(\gammac \le \gamma < \gammai ,t) &= t^4\Fini(\gamma,t) \\ &
+ c (\gamma + \gammad)^{3\nu-4} \left[ \guncs{\dfrac{\gammad}{\gamma +
      \gammad}} - \guncs{\dfrac{\gammad}{\gammai + \gammad}} \right],
\end{aligned}
\end{equation}
and in the domain $\gamma < \gammac$, one gets
\begin{equation}
\begin{aligned}
\label{eq:regcritgc}
t^4 \calF(\gamma < \gammac ,t) &= t^4\Fini(\gamma,t) \nonumber \\ & +
\dfrac{\cc}{\muc} (\gamma + \gammad)^{2 \chic - 3}
\left[\funcc{\dfrac{\gammad}{\gamma + \gammad}} -
  \left(\dfrac{\gamma+\gammad}{\gammac+\gammad}\right)^{\muc}
  \funcc{\dfrac{\gammad}{\gammac+\gammad}} \right]\nonumber \\ & +
c (\gamma + \gammad)^{3\nu-4} \left[\guncs{\dfrac{\gammad}{\gammac+\gammad}}  -
\guncs{\dfrac{\gammad}{\gammai + \gammad}} \right].
\end{aligned}
\end{equation}
The logarithmic growth in time has disappeared, and the solutions are
now scaling. Taking Eq.~\eqref{eq:regcritgd} in the limit $\gamma
\gg \gammad$ and neglecting all terms associated with the initial
conditions, one gets
\begin{equation}
t^4 \calF(\gamma \gg \gammad,t) \simeq c \, \gamma^{3\nu-4}
\ln\left(\dfrac{\gammai}{\gamma} \right).
\end{equation}
The limit $\gammac < \gamma \ll \gammad$ consistently gives
\begin{equation}
t^4 \calF(\gammac <\gamma \ll \gammad,t) = c\, \gammad^{3\nu-4}
\left[\dfrac{1}{3 - 3\nu} \left(\dfrac{\gamma}{\gammad}
  \right)^{3\nu-3} + \ln \left(\dfrac{\gammad}{\gammai} \right)
  \right],
\label{eq:regcritinter}
\end{equation}
and the distribution is back to the scaling power law
$\gamma^{3\nu-3}$.
\\
Finally, small loops with $\gamma \ll \gammac \ll \gammad$ also
scale with a flat distribution as
\begin{equation}
  t^4 \calF(\gamma \ll \gammac,t) = c\, \gammad^{3\nu-4}
  \left[\left(\dfrac{1}{3-3\nu} + \dfrac{1}{2\chic-2} \right) \left(
    \dfrac{\gammac}{\gammad}\right) ^{3\nu-3} +
    \ln{\left(\dfrac{\gammad}{\gammai}\right)} \right].
\label{eq:regcritUV}
\end{equation}

In conclusion, the IR-regularisation we have used solves the
logarithmic time divergence of the loop distribution which now reaches
scaling on all length scales. For $\gamma \ll \gammad$,
Eqs.~\eqref{eq:regcritinter} and \eqref{eq:regcritUV} compared to
Eqs.~\eqref{eq:critinter} and \eqref{eq:critUV} show that the
regularisation is neat, the dependence of the loop distribution with
respect to $\gamma$ is not affected. However, for $\gamma > \gammad$,
the power law behaviour now receives a logarithmic correction. We
therefore conclude that the critical loop production function, even
regularised, exhibits a IR sensitivity.

\subsection{Non-critical loop production function}
\label{sec:regnocritIR}

The calculation follows in all points the one of
section~\ref{sec:regcritIR} and applies to both sub- and
super-critical cases, $\mu > 0$ and $\mu < 0$. The full solution is
presented in the appendix~\ref{app:fullregs} and we focus below on the
asymptotic behaviour only. For the purely IR domain, $\gamma >
\gammai$, the solution is still given by Eq.~\eqref{eq:reggi}, our
IR-regulator assuming an exactly vanishing production function
there. Again neglecting all transients, the solution in the domain
$\gammac \le \gamma < \gammai$ reads
\begin{equation}
\begin{aligned}
  \label{eq:reggd}
t^4 \calF(\gammac \le \gamma < \gammai ,t) &=  t^4\Fini(\gamma,t)
 + \dfrac{c}{\mu}(\gamma +
\gammad)^{2\chi-3} \funcs{\dfrac{\gammad}{\gamma + \gammad}} \\  & -
\dfrac{c}{\mu} (\gamma + \gammad)^{3\nu-4} \left(\gammai +
\gammad\right)^{-\mu} \funcs{\dfrac{\gammad}{\gammai + \gammad}},
\end{aligned}
\end{equation}
while for $\gamma < \gammac$ one obtains
\begin{equation}
\begin{aligned}
\label{eq:reggc}
t^4 \calF(\gammap \le \gamma < \gammac ,t) &=
t^4\Fini(\gamma,t)  \\ & +
\dfrac{\cc}{\muc} (\gamma + \gammad)^{2 \chic - 3}
\left[\funcc{\dfrac{\gammad}{\gamma + \gammad}} -
  \left(\dfrac{\gamma+\gammad}{\gammac+\gammad}\right)^{\muc}
  \funcc{\dfrac{\gammad}{\gammac+\gammad}} \right] \\ & +
\dfrac{c}{\mu} (\gamma + \gammad)^{3\nu-4} \left(\gammac + \gammad
\right)^{-\mu} \funcs{\dfrac{\gammad}{\gammac+\gammad}}
\\ & - \dfrac{c}{\mu} (\gamma + \gammad)^{3\nu-4} \left(\gammad +
\gammai \right)^{-\mu} \funcs{\dfrac{\gammad}{\gammai + \gammad}}.
\end{aligned}
\end{equation}
Here again, the IR cut in the loop production functions can be viewed
as the same formal replacement as \eqref{eq:formal}. Let us now
discuss separately the physical consequences for the sub- and
super-critical loop production functions and we start by the simplest
case which is the sub-critical one.

\subsubsection{Sub-critical case}
\label{sec:regsubcritIR}

Even if sub-critical loop production functions produce a scaling loop
distribution without any regularisation, one may wonder whether
forcing the (unnecessary, for scaling!) cut at $\gamma > \gammai$ can
significantly change the shape of the scaling loop distribution.

At late times, and for sub-critical production functions, $\mu>0$, we
can take the limit $\gamma \gg \gammad$ of \eqref{eq:reggd}
\begin{equation}
t^4 \calF(\gammad \ll \gamma  < \gammai, t) \simeq \dfrac{c}{\mu}
\gamma^{2\chi-3} \left[1 - \left(\dfrac{\gamma}{\gammai}\right)^{\mu}\right].
\end{equation}
Compared to Eq.~\eqref{eq:approxsuperd}, we see that the correction
term $(\gamma/\gammai)^{\mu}$ induced by the IR-regularisation has an
effect only for $\gamma \simeq \gammai$ and becomes rapidly
negligible as soon as $\gamma < \gammai$. For loops having
$\gamma \ll \gammad$, we get
\begin{equation}
t^4 \calF(\gammac \le \gamma  \ll \gammad, t \ge \tc) \simeq \dfrac{c}{2-2\chi}
\dfrac{\gamma^{2\chi-2}}{\gammad} \,,
\end{equation}
the correction $(\gammad/\gammai)^\mu$ can always be safely
ignored. Finally, for loops smaller than the GW backreaction length,
$\gamma \ll \gammac$, we recover Eq.~\eqref{eq:subcritUV}. The
IR-correction added corresponds to the fourth term of
Eq.~\eqref{eq:reggc} and remains again always negligible for $\mu>0$.

We therefore conclude that sub-critical loop production functions
yield scaling loop distributions that are immune to the IR behaviour of
the network.

\subsubsection{Super-critical case}
\label{sec:regsupcritIR}

For super-critical values of $\chi > \chicrit$, we have $\mu < 0$ and most of
the arguments applying for $\mu > 0$ are now reversed. For instance,
the limit $\gammad \ll \gamma < \gammai$ becomes
\begin{equation}
t^4 \calF(\gammad \ll \gamma < \gammai, t)  \simeq  -
\dfrac{c}{\mu} \gamma^{2\chi-3}
\left[\left(\dfrac{\gammai}{\gamma}\right)^{-\mu} - 1\right]  \simeq -
\dfrac{c}{\mu} \gammai^{-\mu} \,  \gamma^{3 \nu -4}.
\end{equation}
The time divergence of the loop distribution is solved but the
power-law exponent has been changed from $2\chi -3$ to $3\nu-4$, see
Eq.~\eqref{eq:pathomuneg}. For smaller loops, we get
\begin{equation}
t^4 \calF(\gammac \le \gamma \ll \gammad, t) \simeq
-\dfrac{c}{\mu}\, \gammad^{2\chi-3} \left[-\dfrac{\mu}{2-2\chi}
  \left(\dfrac{\gamma}{\gammad}\right)^{2\chi-2} +
  \left(\dfrac{\gammai}{\gammad} \right)^{-\mu}\right].
\label{eq:regsupgdapprox}
\end{equation}
Since $\gammai/\gammad \gg 1$, the IR cut is adversely
introducing a new length scale! Thus, let us define $\gammair$ by
\begin{equation}
\gammair \equiv \left[\dfrac{-\mu}{(2-2\chi)\gammai^{-\mu}}
\right]^{\frac{1}{2-2\chi}} \gammad^{\frac{3-3\nu}{2-2\chi}}.
\label{eq:gammair}
\end{equation}
For $\gamma > \gammair$, Eq.~\eqref{eq:regsupgdapprox} shows that the
loop distribution is flat, the dependence in $\gamma$ remains
negligible compared to the constant term introduced by the
regularisation. On the contrary, for $\gamma < \gammair$, we recover a
power-law behaviour as $\gamma^{2\chi-2}$. This new IR scale is
relevant only if $\gammair > \gammac$, which is model- and
regularisation-dependent. Nonetheless, if we assume the dependency in
$\GU$ for $\gammad$ given in Eq.~\eqref{eq:gammad},
\begin{equation}
  \gammair  \propto(\GU)^{\frac{3-3\nu}{2-2\chi}},
\end{equation}
and using Eq.~\eqref{eq:gammac}
\begin{equation}
\dfrac{\gammair}{\gammac} \propto (\GU)^{\frac{4 \chi^2 - 2\chi+1-3\nu}{2-2\chi}}.
\label{eq:gammairvsc}
\end{equation}
This defines a particular value for $\chi$, namely
\begin{equation}
\chiir \equiv \dfrac{1 + \sqrt{12\nu-3}}{4}\,,
\end{equation}
whose numerical value in the radiation era is $\chiir \simeq 0.683$
and $\chiir \simeq 0.809$ for the matter era. For all values $\chicrit
<\chi < \chiir$, the exponent of Eq.~\eqref{eq:gammairvsc} is
negative. For $\GU$ small enough, we generically have $\gammair >
\gammac$. As a result, the regularised loop distribution is now
scaling but exhibits a new plateau for $\gammair < \gamma < \gammad$,
which smoothly connects to the $\gamma^{2\chi-2}$ behaviour in the
domain $\gammac \le \gamma < \gammair$. For larger values of $\chi >
\chiir$ (and deeper negative values of $\mu$), only the plateau exists
in the whole domain $\gammac \le \gamma <\gammad$, the amplitude of
the constant term $(\gammai/\gammad)^{-\mu}$ is so large that it
erases any features that could be associated with the scale of
gravitational wave emission. This situation is actually reminiscent
with the time-divergent behaviour discussed in
section~\ref{sec:supercrit}.

Finally, for the very small loops, $\gamma \ll \gammac$, with $\gammac
\ll \gammad$, the loop distribution reads
\begin{equation}
t^4 \calF(\gamma  \ll  \gammac, t \ge \tc) \simeq c
\left(\dfrac{1}{2-2\chi} + \dfrac{1}{2\chic-2} \right)
\dfrac{\gammac^{2\chi-2}}{\gammad} -\dfrac{c}{\mu} \gammai^{-\mu}
\gammad^{3\nu-4} + \order{\gammad^{2\chi-3}}.
\label{eq:regsupgcapprox}
\end{equation}
It is scaling with a plateau behaviour. The amplitude of the plateau
is either given by the first term, the one varying as
$\gammac^{2\chi-2}/\gammad$, or the second term which is proportional
to $\gammai^{-\mu} \gammad^{3\nu-4}$. That depends on their relative
amplitude. Neglecting the terms in $\chic$, which are sub-dominant,
the ratio $\calR$ of the first to second term in the right-hand side
of Eq.~\eqref{eq:regsupgcapprox} simplifies to
\begin{equation}
\calR = \left(\dfrac{\gammair}{\gammac} \right)^{2-2\chi}.
\end{equation}
Consistently with the behaviour in the $\gamma > \gammac$ domains, for
$\chicrit <\chi < \chiir$, one always has $\calR \gg 1$ and the
regularization effects are small. Only for $\chi > \chiir$, the
plateau at $\gamma < \gammac$ is dominated by the regulator and
continuously matches the one at $\gamma > \gammac$.

We conclude that IR-regularisation of super-critical loop production
functions solves their time-divergence, but this has the
consequence of significantly modifying the shape of the actual scaling
distribution. The results are therefore strongly IR-sensitive.

\subsection{Influence of a power-law IR-regularisation}
\label{sec:smoothIR}

Considering the strong dependence of the loop number density on the
parameter $\gammai$, one might ask whether the shape of the IR-cutoff
has an additional influence on the results. To perform this analysis,
we introduce an additional source term $\ci \gamma^{2\chii-3}
\heaviside{\gamma - \gammai}$ to the collision term of the Boltzmann
equation~\eqref{eq:Pbig} and, neglecting all possible transients,
compute its contribution, say $t^4\calFi$, to the asymptotic loop
number density. For this source term to be a well-behaved
IR-regulator, it has to fulfil two conditions. First $\mui > 0$
otherwise we expect this term to present the same time-divergent
behaviour as the critical and super-critical distributions. Then, we
should have $c_\infty = c \gammai^{2(\chi-\chii)}$ for the loop
production function to be continuous in \(\gammai\). Then the
contribution of such a power-law cutoff is
\begin{equation}
\begin{aligned}
    \label{eq:plreg}
    t^4\calFi(\gamma<\gammai) &=
    \frac{\ci}{\mui} \frac{(\gamma + \gammad)^{3\nu-4}}{(\gammai +
      \gammad)^{\mui}} \funci{\frac{\gammad}{\gamma_\infty + \gammad}} \\
    & - \frac{\ci}{\mui} (\gamma+\gammad)^{2\chii-3}
    \left(\frac{\tini}{t}\right)^{\mui}
    \funci{\frac{\gammad}{\gamma+\gammad}\frac{\tini}{t}},
\end{aligned}
\end{equation}
where
\begin{equation}
    \funci{x} \equiv \hypergauss{3-2\chii}{\mui}{\mui+1}{x}.
\end{equation}
The condition $\mui>0$ ensures that all time-dependent contributions
are suppressed at late-times.  Under the assumption that \(\gammad \ll
\gammai\), the contribution to the scaling loop number density coming
from the power-law cutoff is
\begin{equation}
t^4\calFi(\gamma<\gammai) = \ci \frac{(\gamma+\gammad)^{3\nu-4}}{\mui
  \gammai^{\mui}} = c \frac{(\gamma+\gammad)^{3\nu-4}}{\mui
  \gammai^{\mu}} \,.
\end{equation}
This additional part generically contributes and can modify the shape
of the loop distribution, as for instance it would modify the value of
$\gammair$ for the super-critical case in
Eq.~\eqref{eq:gammair}. However, for large enough values of $\mui$,
namely for $\mui \gg |\mu|$, it can safely be neglected with respect
to the one computed earlier. As a result, the IR-regularisation
effects we have found in the previous section are relatively generic
in the sense that they are not simply induced by the choice of an
infinitely sharp cut in the LPF but rather by suppressing the
production of large loops.

\section{Conclusions}

\begin{table}
\begin{center}
\renewcommand{\arraystretch}{2}
\begin{tabular}{|c|c|c|c|c|c|c|c|}
  \hline Type & \(\gamma < \gammac\) & \(\gammac < \gamma < \gammair\)
  & \(\gammair < \gamma < \gammad\) & \(\gamma > \gammad\)\\ \hline
  \multirow{1}{\roww}{Sub-critical}
  &\(\dfrac{c}{2-2\chi}\gammac^{2\chi-2}\gammad^{-1}\)&\(\dfrac{c}{2-2\chi}
  \gamma^{2\chi-2}\gammad^{-1}\)&\(-\)&\(\dfrac{c}{\mu}\gamma^{2\chi-3}\)
  \\ \multirow{1}{\roww}{IR Critical}
  &\(\dfrac{c}{3-3\nu}\gammac^{3\nu-3}\gammad^{-1}\)&\(\dfrac{c}{3-3\nu}\gamma^{3\nu-3}
  \gammad^{-1}\)&\(-\)&\(c\gamma^{3\nu-4}\ln\left(\dfrac{\gammai}{\gamma}\right)\)
  \\ \multirow{1}{\roww}{IR Super-critical with $\chi <
    \chiir$}&\(\dfrac{c}{2-2 \chi}\gammac^{2\chi-2}\gammad^{-1}\)&\(\dfrac{c}{2-2\chi}
  \gamma^{2\chi-2}\gammad^{-1}\)&\(-\dfrac{c}{\mu}\gammai^{-\mu}\gammad^{3\nu-4}\)&\(-\dfrac{c}{\mu}
  \gammai^{-\mu}\gamma^{3\nu-4}\)
  \\ \multirow{1}{\roww}{IR Super-critical with $\chi >
    \chiir$}&\(-\dfrac{c}{\mu}\gammai^{-\mu}\gammad^{3\nu-4}\)&\(-\dfrac{c}{\mu}
  \gammai^{-\mu}\gammad^{3\nu-4}\)&\(-\dfrac{c}{\mu}\gammai^{-\mu}\gammad^{3\nu-4}\)
  &\(-\dfrac{c}{\mu}\gammai^{-\mu}\gamma^{3\nu-4}\)
  \\ \hline
\end{tabular}
\caption{Asymptotic contributions to the loop number density assuming
  a ``strong'' enough infrared cutoff. With this assumption, both
  critical and super-critical loop number densities scale with time
  but their shape is modified compared to the unregularised ones (see
  Table~\ref{table:dominantnoircutoff}).}
\label{table:dominantircutoff}
\end{center}
\end{table}

\begin{figure}
  \begin{center}
    \includegraphics[width=\onefigw]{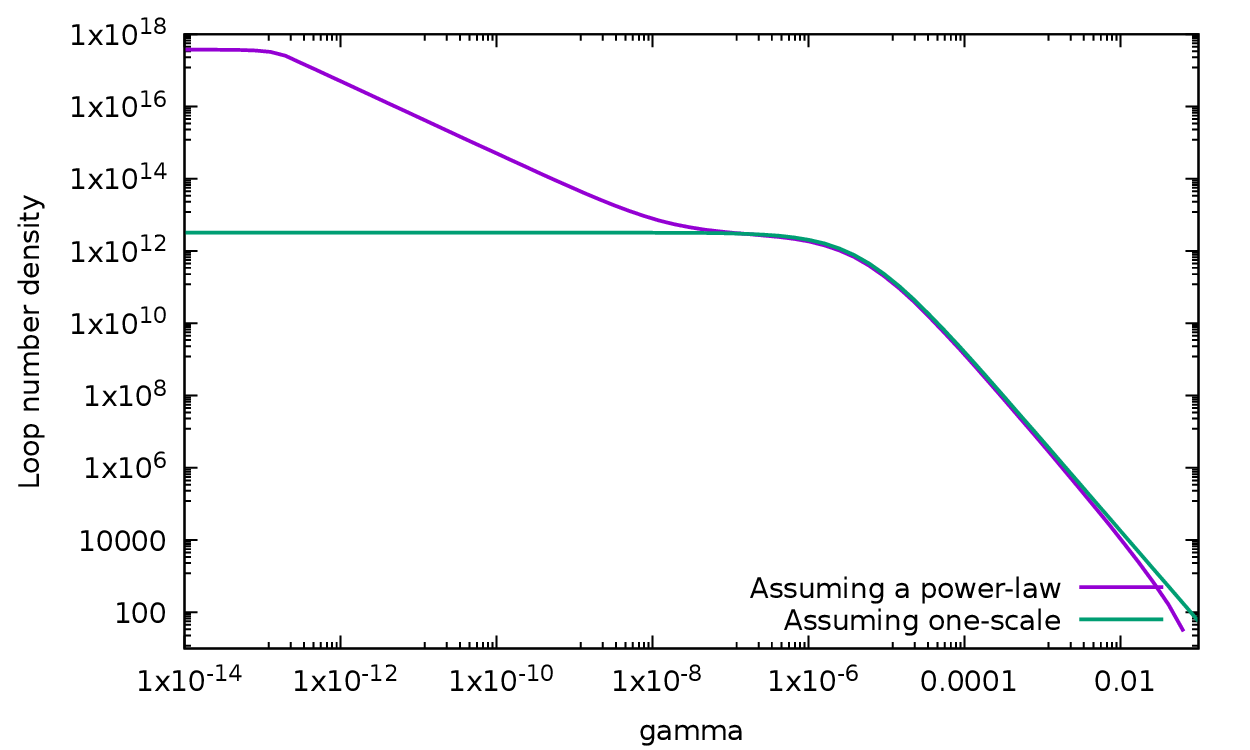}
    \caption{Difference between loop distributions in the radiation
      era generated by a Dirac distribution LPF (green lower curve)
      and a super-critical, IR-regularised, Polchinski-Rocha one
      (purple top curve). Given a super-critical power-law loop
      production function, one can reproduce the large scale behavior
      of the loop distribution with a Dirac distribution for the loop
      production function (see section \ref{sec:deltaf}).  Doing so,
      one loses the small-scale behavior of the loop distribution.
      For illustration purposes, we have chosen $GU=10^{-7}$,
      $c\simeq0.25$ and $\gammai=0.1$ for the super-critical LPF and
      $c \simeq 5.7$ for the Dirac distribution.}
    \label{fig:dd}
  \end{center}
\end{figure}

The aim of this paper has been to carry out an exhaustive study of the
effect of the loop production function on the cosmological
distribution of loops. As explained in the Introduction, numerical
simulations of Nambu-Goto cosmic string networks are not currently
able to capture some important physical effects at very small scales,
for instance GW emission and its backreaction effects. Hence
determining the loop distribution, by construction, requires an
interplay between numerical results (valid for larger loops where the
extra physics should be negligible) and analytical modelling.

The analytical tool used to solve for the loop distribution is the
Boltzmann equation \eqref{eq:Pbig}. On the one hand, we have shown
that very different LPF, namely, a Dirac distribution motivated by the
one-scale model, and a sub-critical Polchinski-Rocha power-law
distribution ($\chi < \chicrit$) taking into account the small-scale
structure built up on the strings, can give rise to a scaling,
power-law, distribution on large scales, albeit with different
power-law exponents. On the other hand, we have found that the actual
value of the power-law exponent, i.e., the value of $\chi$ with
respect to $\chicrit=(3\nu-1)/2$, produces very different
behaviours. Critical and super-critical LPFs ($\chi \ge \chicrit$)
lead to time-divergent loop distributions, which do not scale. The
critical case however exhibits only a logarithmic growth for large
loops, $\gamma \ge \gammad$, and a very long transient scaling for the
smaller ones, $\gamma < \gammad$, that can last longer than the age of
the universe.

The divergent behaviour of the critical and super-critical cases has
been traced back to a relative over-production of large loops with
respect to small loops and we have shown that it can be regularised by
arbitrarily assuming that the PR loop production function vanishes
above some length scale $\gammai$. We find, however, that although
such a IR regularisation fixes the time divergence, it is also
changing the shape of the loop distribution. For this reason, we
conclude that both the critical and super-critical LPF are genuinely
IR-sensitive. For the critical case, we find that the large loop
distribution acquires a new logarithmic dependence in $\gamma$ (again
for $\gamma \ge \gammad$).  On the small scales $\gamma < \gammad$,
the predictions are all very different and depend on both the PR
exponent $\chi$ and on the IR regulator. The results of our study are
summarised in Tables~\ref{table:dominantnoircutoff} and
\ref{table:dominantircutoff}, where we give the asymptotic
contributions to the loop number density on all scales \(\gamma\)
depending on the value of the parameter \(\mu\equiv 3\nu-2\chi-1\) (it
vanishes for $\chi=\chicrit$). Let us notice that for extreme values
of $\GU$, and times close to the transition from the radiation to the
matter era, these results may not apply and one should rely on the
complete solutions given in the appendices.

It is interesting to observe from the last row of
Table~\ref{table:dominantircutoff}, that in the super-critical case
only and assuming an IR cutoff, the obtained distribution for large
$\gamma \ge \gammad$ is essentially identical to that obtained from
assuming a Dirac distribution for the LPF. In particular, for large
$\gamma$ there is a $-5/2$ power-law in the radiation era and $-2$
power-law in the matter era, which are the values for the exponents
that we have obtained in Section~\ref{sec:deltaf}. At the same time,
both distributions are completely different on smaller scales. This is
illustrated in figure~\ref{fig:dd}.

In this paper, following Ref.~\cite{Lorenz:2010sm}, we have also
introduced a small distance scale $\gammac$ below which gravitational
backreaction is expected to be important. Generically, for $\gammac
\ll \gammad$, and for all values of $\chi$, the amplitude of the loop
distribution at small $\gamma < \gammac$ is enhanced relative to the
Dirac distribution LPF, and, as discussed in
Ref.~\cite{Ringeval:2017eww}, this leads to observational consequences
on the SGWB. Another interesting feature we have not discussed in the
main text concern the various transient domains associated with the IR
regularisation. They are excited soon after the network is created,
but also during the transition from the radiation to matter era. As
such, they may also lead to interesting phenomenological consequences,
in particular regarding a gravitational wave signature.

\acknowledgments

It is a pleasure to thank J.~J. Blanco-Pillado, K.~Olum and
J.~Wachter for motivating discussions, as well as the participants and
organisers of the 2018 Lorentz Center Cosmic Topological Defects
Workshop in Leiden (Netherlands).  We are also grateful to the LISA
cosmic string group for stimulating discussions related to cosmic
strings. The work of C.\ R. is supported by the ``Fonds de la
Recherche Scientifique - FNRS'' under Grant
$\mathrm{N^{\circ}T}.0198.19$. The work of M.\ S. is supported in part
by the Science and Technology Facility Council (STFC), United Kingdom,
under the research grant ST/P000258/1.

\appendix

\section{Complete solutions}
\label{app:fullsols}

In this appendix, we give the explicit expressions of the solution of
the Boltzmann equation~\eqref{eq:evolgam}. Details of the calculation
can be found in Refs.~\cite{Lorenz:2010sm, Peter:2013jj} and we here
simply report the results.

\subsection{Non-critical loop production function}

\label{sec:fullnc}

For the piecewise PR loop production function given in
Eqs.~\eqref{eq:Pbig} and \eqref{eq:Psmall}, assuming $\chi \ne
\chicrit$, one gets
\begin{align}
\label{eq:fullsolgd}
t^4 \calF(\gamma \ge \gammac,t) &= \left(\dfrac{t}{\tini} \right)^4
\left(\dfrac{\aini}{a}\right)^3 \tini^4\,
\calNini\negthickspace\left\{\left[\gamma + \gammad\left(1 -
  \dfrac{\tini}{t} \right) \right] t \right\} \nonumber \\ & +
\dfrac{c}{\mu}(\gamma + \gammad)^{2\chi-3}
\funcs{\dfrac{\gammad}{\gamma + \gammad}} \nonumber \\ & -
\dfrac{c}{\mu} (\gamma + \gammad)^{2\chi - 3} \left(\dfrac{t}{\tini}
\right)^{-\mu} \funcs{\dfrac{\gammad}{\gamma + \gammad}
  \dfrac{\tini}{t}}, \\
\label{eq:fullsolgc}
t^4\calF(\gammax \le \gamma < \gammac,t) &= \left(\dfrac{t}{\tini}
\right)^4 \left(\dfrac{\aini}{a}\right)^3 \tini^4\,
\calNini\negthickspace\left\{\left[\gamma + \gammad\left(1 -
  \dfrac{\tini}{t} \right) \right] t \right\} \nonumber \\ & +
\dfrac{\cc}{\muc} (\gamma + \gammad)^{2 \chic - 3}
\left[\funcc{\dfrac{\gammad}{\gamma + \gammad}} -
  \left(\dfrac{\gamma+\gammad}{\gammac+\gammad}\right)^{\muc}
  \funcc{\dfrac{\gammad}{\gammac+\gammad}} \right]\nonumber \\ & +
\dfrac{c}{\mu} (\gamma + \gammad)^{3 \nu - 4}
\left(\gammac+\gammad \right)^{-\mu}
\funcs{\dfrac{\gammad}{\gammac+\gammad}} \nonumber \\ & -
\dfrac{c}{\mu} (\gamma + \gammad)^{2\chi - 3} \left(\dfrac{t}{\tini}
\right)^{-\mu} \funcs{\dfrac{\gammad}{\gamma + \gammad} \dfrac{\tini}{t}}, \\
\label{eq:fullsolgo}
t^4 \calF(0<\gamma < \gammax,t) &= \left(\dfrac{t}{\tini} \right)^4
\left(\dfrac{\aini}{a}\right)^3 \tini^4\,
\calNini\negthickspace\left\{\left[\gamma + \gammad\left(1 -
  \dfrac{\tini}{t} \right) \right] t \right\}\nonumber \\ & +
\dfrac{\cc}{\muc} (\gamma + \gammad)^{2 \chic - 3}
\funcc{\dfrac{\gammad}{\gamma + \gammad}} \nonumber \\ & -
\dfrac{\cc}{\muc} (\gamma + \gammad)^{2\chic - 3}
\left(\dfrac{t}{\tini} \right)^{-\muc} \funcc{\dfrac{\gammad}{\gamma +
    \gammad} \dfrac{\tini}{t}}.
\end{align}
where we recap that
\begin{equation}
\funcs{x} \equiv \hypergauss{3-2\chi}{\mu}{\mu+1}{x}, \qquad \funcc{x}
\equiv \hypergauss{3-2\chic}{\muc}{\muc+1}{x}.
\end{equation}
and
\begin{equation}
\mu \equiv 3 \nu - 2 \chi -1, \qquad \muc \equiv 3 \nu - 2 \chic -1.
\end{equation}
There is a transient domain for loops having $\gamma$ smaller than
\begin{equation}
\label{eq:gammax}
\gammax(t) \equiv (\gammac + \gammad) \dfrac{\tini}{t} - \gammad,
\end{equation}
which describes a virgin population of loops that started their
evolution with a $\gamma < \gammac$ and which have never been
contaminated by shrunk loops produced at $\gamma > \gammac$. This
population of loops cannot exist forever and the domain disappears for
times $t \ge \tx$ where $\gammax(\tx) = 0$.

\subsection{Critical loop production function}
\label{sec:fullc}

In the critical case, the piecewise loop production function is given
by Eq.~\eqref{eq:Pbig} in the domain $\gamma \ge \gammac$ with
$\chi=\chicrit$, and Eq.~\eqref{eq:Psmall} for $\gamma < \gammac$
which is unchanged. The solution reads
\begin{align}
\label{eq:fullcritgd}
t^4 \calF(\gamma \ge \gammac,t) &= \left(\dfrac{t}{\tini} \right)^4
\left(\dfrac{\aini}{a}\right)^3 \tini^4\,
\calNini\negthickspace\left\{\left[\gamma + \gammad\left(1 -
  \dfrac{\tini}{t} \right) \right] t \right\} \nonumber \\ & + c
(\gamma + \gammad)^{3\nu-4} \left[ \guncs{\dfrac{\gammad}{\gamma + \gammad}}
- \guncs{\dfrac{\gammad}{\gamma + \gammad} \dfrac{\tini}{t}} \right],
\\
\label{eq:fullcritgc}
t^4\calF(\gammax \le \gamma < \gammac,t) &= \left(\dfrac{t}{\tini}
\right)^4 \left(\dfrac{\aini}{a}\right)^3 \tini^4\,
\calNini\negthickspace\left\{\left[\gamma + \gammad\left(1 -
  \dfrac{\tini}{t} \right) \right] t \right\} \nonumber \\ & +
\dfrac{\cc}{\muc} (\gamma + \gammad)^{2 \chic - 3}
\left[\funcc{\dfrac{\gammad}{\gamma + \gammad}} -
  \left(\dfrac{\gamma+\gammad}{\gammac+\gammad}\right)^{\muc}
  \funcc{\dfrac{\gammad}{\gammac+\gammad}} \right]\nonumber \\ & +
c (\gamma + \gammad)^{3\nu-4} \left[\guncs{\dfrac{\gammad}{\gammac+\gammad}}  -
\guncs{\dfrac{\gammad}{\gamma + \gammad} \dfrac{\tini}{t}} \right],  \\
\label{eq:fullcritgo}
t^4 \calF(0<\gamma < \gammax,t) &= \left(\dfrac{t}{\tini} \right)^4
\left(\dfrac{\aini}{a}\right)^3 \tini^4\,
\calNini\negthickspace\left\{\left[\gamma + \gammad\left(1 -
  \dfrac{\tini}{t} \right) \right] t \right\}\nonumber \\ & +
\dfrac{\cc}{\muc} (\gamma + \gammad)^{2 \chic - 3}
\funcc{\dfrac{\gammad}{\gamma + \gammad}} \nonumber \\ & -
\dfrac{\cc}{\muc} (\gamma + \gammad)^{2\chic - 3}
\left(\dfrac{t}{\tini} \right)^{-\muc} \funcc{\dfrac{\gammad}{\gamma +
    \gammad} \dfrac{\tini}{t}},
\end{align}
where we recap that the first integral $\guncs{x}$ is given by
\begin{equation}
\begin{aligned}
  \guncrad{x} \equiv \ln\left(\dfrac{1-\sqrt{1-x}}{1+\sqrt{1-x}}\right)
+ \dfrac{2}{3} \dfrac{4-3 x}{(1-x)^{3/2}}\,,\qquad
\guncmat{x}  \equiv \dfrac{1}{1-x} \ln\left(\dfrac{1-x}{x}\right),
\end{aligned}
\end{equation}
in the radiation and matter era, respectively. Notice that the small
scales transient, Eq.~\eqref{eq:fullcritgo}, is identical to
Eq.~\eqref{eq:fullsolgo}. To ease comparison with the non-critical
case, let us stress that for $\chi=\chicrit$, one has $\mu=0$ and
$2\chic-3 = 3\nu -4$ such that the critical functional shape is
smoothly interpolating between the sub- and super-critical solutions
presented in section~\ref{sec:fullnc}.

\section{Sharp infrared regularisation}
\label{app:fullregs}

\begin{figure}
    \centering
    \includegraphics[width=\onefigw]{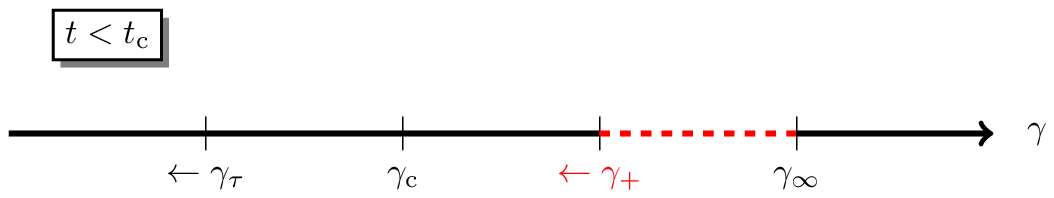}
    \includegraphics[width=\onefigw]{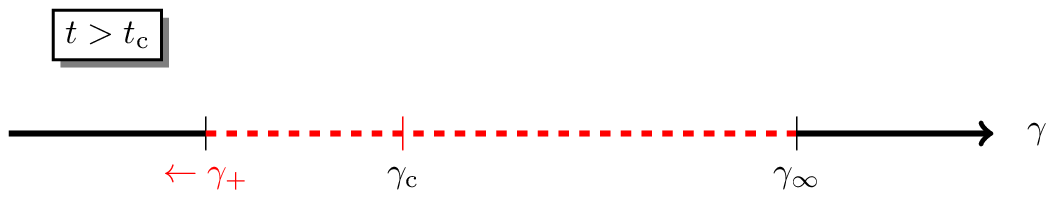}
    \caption{Schematic representation of the different domains of
      $\gamma$ for $t<\tc$ and for $t > \tc$. The black regions are
      causally disconnected from the cutoff at $\gammai$ such that the
      solutions are exactly the same as the non-regularised ones. On
      the contrary, this is not the case in the red dotted regions and
      one has to use the modified expression for $t^4\calF(\gamma \ge
      \gammap,t)$ (see text).}
    \label{fig:clarity}
\end{figure}

The sharp IR-regularisation consists in cutting the loop production
function above some length scale $\gammai$. Therefore, it is a
piecewise function over three domains: for $\gamma < \gammac$ it is
given by Eq.~\eqref{eq:Psmall}, for $\gammac \le \gamma < \gammai$ by
Eq.~\eqref{eq:Pbig} and for $\gamma \ge \gammai$ it is vanishing. The
new length scale $\gammai$ introduces a new, time-dependent, length
scale defined by
\begin{equation}
\gammap(t) \equiv \left(\gammad + \gammai\right) \dfrac{\tini}{t} - \gammad.
\label{eq:gammap}
\end{equation}
Physically its meaning is the following: if we consider a loop which
was created at time $\tini$ with the maximal possible size $\gammai
\tini$, then at time $t$ its length is $\ellp =\gammap t$. Therefore,
at time $t$, loops having $\gamma < \gammap(t)$ are not affected by
the IR cutoff and the non-regularised solutions are still valid. On
the contrary, the loop distribution for $\gamma > \gammap(t)$ has to
be re-derived by solving the Boltzmann equation and satisfying the two
continuity conditions at $\gamma=\gammac$ and $\gamma=\gammai$. In
doing so, we must distinguish the cases for which $\gammap(t) >
\gammac$ from those having $\gammap(t) < \gammac$. To this end, we
define $t=\tc$ through $\gammap(\tc)=\gammac$ from which
\begin{equation}
\tc \equiv \dfrac{\gammai + \gammad}{\gammad + \gammac} \tini.
\label{eq:tc}
\end{equation}
If we compare Eqs.~\eqref{eq:gammax} and \eqref{eq:gammap}, we have
$\gammax(\tini)=\gammac$ and $\gammap(\tini)=\gammai$; the domains
never collide: $\gammap(t) - \gammax(t) = (\gammai - \gammac)(\tini/t)
> 0$. At last, the domain $\gamma < \gammap(t)$ disappears completely
for $t>\tp$ where
\begin{equation}
\tp \equiv \left(1+\dfrac{\gammai}{\gammad} \right) \tini,
\end{equation}
which is defined by $\gammap(\tp)=0$. The different transient domains
thus defined are summarized in figure~\ref{fig:clarity}. In practice,
the solution is affected by the IR cutoff only within the red dashed
zones appearing in this figure, but for completeness, we give, and
repeat, the solutions in all contiguous domains.

\subsection{Non-critical loop production function}

We distinguish the two cases, $t \le \tc$ and $t>\tc$. During the
relaxation period $t\le\tc$, the solution reads
\begingroup
  \allowdisplaybreaks
\begin{align}
  \label{eq:regrelaxgi}
  t^4 \calF(\gamma \ge \gammai ,t < \tc ) &=  \left(\dfrac{t}{\tini} \right)^4
  \left(\dfrac{\aini}{a}\right)^3 \tini^4\,
  \calNini\negthickspace\left\{\left[\gamma + \gammad\left(1 -
    \dfrac{\tini}{t} \right) \right] t \right\}, \\
  \label{eq:regrelaxgd}
  t^4 \calF(\gammap \le \gamma < \gammai ,t < \tc ) &=  \left(\dfrac{t}{\tini} \right)^4
  \left(\dfrac{\aini}{a}\right)^3 \tini^4\,
  \calNini\negthickspace\left\{\left[\gamma + \gammad\left(1 -
    \dfrac{\tini}{t} \right) \right] t \right\} \nonumber \\   & + \dfrac{c}{\mu}(\gamma +
  \gammad)^{2\chi-3} \funcs{\dfrac{\gammad}{\gamma + \gammad}}  \nonumber \\  & -
  \dfrac{c}{\mu} (\gamma + \gammad)^{3\nu-4} \left(\gammai +
  \gammad\right)^{-\mu} \funcs{\dfrac{\gammad}{\gammai + \gammad}},
 \\
  \label{eq:regrelaxgp}
  t^4 \calF(\gammac \le \gamma < \gammap ,t < \tc) &=
  \left(\dfrac{t}{\tini} \right)^4 \left(\dfrac{\aini}{a}\right)^3
  \tini^4\, \calNini\negthickspace\left\{\left[\gamma + \gammad\left(1 -
    \dfrac{\tini}{t} \right) \right] t \right\} \nonumber \\ & +
  \dfrac{c}{\mu}(\gamma + \gammad)^{2\chi-3}
  \funcs{\dfrac{\gammad}{\gamma + \gammad}} \nonumber \\ & -
  \dfrac{c}{\mu} (\gamma + \gammad)^{2\chi - 3} \left(\dfrac{t}{\tini}
  \right)^{-\mu} \funcs{\dfrac{\gammad}{\gamma + \gammad}
    \dfrac{\tini}{t}},  \\
  \label{eq:regrelaxgc}
  t^4\calF(\gammax \le \gamma < \gammac,t < \tc) &= \left(\dfrac{t}{\tini}
\right)^4 \left(\dfrac{\aini}{a}\right)^3 \tini^4\,
\calNini\negthickspace\left\{\left[\gamma + \gammad\left(1 -
  \dfrac{\tini}{t} \right) \right] t \right\} \nonumber \\ & +
\dfrac{\cc}{\muc} (\gamma + \gammad)^{2 \chic - 3}
\left[\funcc{\dfrac{\gammad}{\gamma + \gammad}} -
  \left(\dfrac{\gamma+\gammad}{\gammac+\gammad}\right)^{\muc}
  \funcc{\dfrac{\gammad}{\gammac+\gammad}} \right]\nonumber \\ & +
\dfrac{c}{\mu} (\gamma + \gammad)^{3 \nu - 4}
\left(\gammac+\gammad \right)^{-\mu}
\funcs{\dfrac{\gammad}{\gammac+\gammad}} \nonumber \\ & -
\dfrac{c}{\mu} (\gamma + \gammad)^{2\chi - 3} \left(\dfrac{t}{\tini}
\right)^{-\mu} \funcs{\dfrac{\gammad}{\gamma + \gammad}
  \dfrac{\tini}{t}}, \\
\label{eq:regrelaxgo}
t^4 \calF(0<\gamma < \gammax,t < \tc) &= \left(\dfrac{t}{\tini} \right)^4
\left(\dfrac{\aini}{a}\right)^3 \tini^4\,
\calNini\negthickspace\left\{\left[\gamma + \gammad\left(1 -
  \dfrac{\tini}{t} \right) \right] t \right\}\nonumber \\ & +
\dfrac{\cc}{\muc} (\gamma + \gammad)^{2 \chic - 3}
\funcc{\dfrac{\gammad}{\gamma + \gammad}} \nonumber \\ & -
\dfrac{\cc}{\muc} (\gamma + \gammad)^{2\chic - 3}
\left(\dfrac{t}{\tini} \right)^{-\muc} \funcc{\dfrac{\gammad}{\gamma +
    \gammad} \dfrac{\tini}{t}}.
\end{align}
\endgroup

For later times, $t \ge \tc$, we get the solution
\begingroup
  \allowdisplaybreaks
\begin{align}
\label{eq:altreggi}
  t^4 \calF(\gamma \ge \gammai ,t \ge \tc ) &=  \left(\dfrac{t}{\tini} \right)^4
  \left(\dfrac{\aini}{a}\right)^3 \tini^4\,
  \calNini\negthickspace\left\{\left[\gamma + \gammad\left(1 -
    \dfrac{\tini}{t} \right) \right] t \right\}, \\
  \label{eq:altreggd}
t^4 \calF(\gammac \le \gamma < \gammai ,t \ge \tc ) &=  \left(\dfrac{t}{\tini} \right)^4
\left(\dfrac{\aini}{a}\right)^3 \tini^4\,
\calNini\negthickspace\left\{\left[\gamma + \gammad\left(1 -
      \dfrac{\tini}{t} \right) \right] t \right\} \nonumber \\   & + \dfrac{c}{\mu}(\gamma +
\gammad)^{2\chi-3} \funcs{\dfrac{\gammad}{\gamma + \gammad}}  \nonumber \\  & -
\dfrac{c}{\mu} (\gamma + \gammad)^{3\nu-4} \left(\gammai +
\gammad\right)^{-\mu} \funcs{\dfrac{\gammad}{\gammai + \gammad}},
 \\
\label{eq:altreggc}
t^4 \calF(\gammap \le \gamma < \gammac ,t \ge \tc) &=
\left(\dfrac{t}{\tini} \right)^4 \left(\dfrac{\aini}{a}\right)^3
\tini^4\, \calNini\negthickspace\left\{\left[\gamma + \gammad\left(1 -
  \dfrac{\tini}{t} \right) \right] t \right\} \nonumber \\ & +
\dfrac{\cc}{\muc} (\gamma + \gammad)^{2 \chic - 3}
\left[\funcc{\dfrac{\gammad}{\gamma + \gammad}} -
  \left(\dfrac{\gamma+\gammad}{\gammac+\gammad}\right)^{\muc}
  \funcc{\dfrac{\gammad}{\gammac+\gammad}} \right]\nonumber \\ & +
\dfrac{c}{\mu} (\gamma + \gammad)^{3\nu-4} \left(\gammac + \gammad
\right)^{-\mu} \funcs{\dfrac{\gammad}{\gammac+\gammad}} \nonumber
\\\ & - \dfrac{c}{\mu} (\gamma + \gammad)^{3\nu-4} \left(\gammad +
\gammai \right)^{-\mu} \funcs{\dfrac{\gammad}{\gammai + \gammad}},\\
\label{eq:altreggp}
  t^4\calF(\gammax \le \gamma < \gammap,t\ge\tc) &= \left(\dfrac{t}{\tini}
\right)^4 \left(\dfrac{\aini}{a}\right)^3 \tini^4\,
\calNini\negthickspace\left\{\left[\gamma + \gammad\left(1 -
  \dfrac{\tini}{t} \right) \right] t \right\} \nonumber \\ & +
\dfrac{\cc}{\muc} (\gamma + \gammad)^{2 \chic - 3}
\left[\funcc{\dfrac{\gammad}{\gamma + \gammad}} -
  \left(\dfrac{\gamma+\gammad}{\gammac+\gammad}\right)^{\muc}
  \funcc{\dfrac{\gammad}{\gammac+\gammad}} \right]\nonumber \\ & +
\dfrac{c}{\mu} (\gamma + \gammad)^{3 \nu - 4}
\left(\gammac+\gammad \right)^{-\mu}
\funcs{\dfrac{\gammad}{\gammac+\gammad}} \nonumber \\ & -
\dfrac{c}{\mu} (\gamma + \gammad)^{2\chi - 3} \left(\dfrac{t}{\tini}
\right)^{-\mu} \funcs{\dfrac{\gammad}{\gamma + \gammad}
  \dfrac{\tini}{t}}, \\
\label{eq:altreggo}
t^4 \calF(0<\gamma < \gammax,t\ge\tc) &= \left(\dfrac{t}{\tini} \right)^4
\left(\dfrac{\aini}{a}\right)^3 \tini^4\,
\calNini\negthickspace\left\{\left[\gamma + \gammad\left(1 -
  \dfrac{\tini}{t} \right) \right] t \right\}\nonumber \\ & +
\dfrac{\cc}{\muc} (\gamma + \gammad)^{2 \chic - 3}
\funcc{\dfrac{\gammad}{\gamma + \gammad}} \nonumber \\ & -
\dfrac{\cc}{\muc} (\gamma + \gammad)^{2\chic - 3}
\left(\dfrac{t}{\tini} \right)^{-\muc} \funcc{\dfrac{\gammad}{\gamma +
    \gammad} \dfrac{\tini}{t}}.
\end{align}
\endgroup

Neglecting all transients and initial condition effects, these equations
show that the IR cut can be viewed as the formal replacement written
in Eq.~\eqref{eq:formal}.

\subsection{Critical loop production function}

For the critical case $\chi=\chicrit$ and the sharp IR cut at
$\gammai$, one gets during the relaxation times $t < \tc$
\begingroup
\allowdisplaybreaks
\begin{align}
\label{eq:regrelaxcritgi}
  t^4 \calF(\gamma \ge \gammai ,t < \tc ) &=  \left(\dfrac{t}{\tini} \right)^4
  \left(\dfrac{\aini}{a}\right)^3 \tini^4\,
  \calNini\negthickspace\left\{\left[\gamma + \gammad\left(1 -
    \dfrac{\tini}{t} \right) \right] t \right\}, \\
  \label{eq:regrelaxcritgd}
t^4 \calF(\gammap \le \gamma < \gammai ,t < \tc ) &= \left(\dfrac{t}{\tini} \right)^4
\left(\dfrac{\aini}{a}\right)^3 \tini^4\,
\calNini\negthickspace\left\{\left[\gamma + \gammad\left(1 -
  \dfrac{\tini}{t} \right) \right] t \right\} \nonumber \\ & + c
(\gamma + \gammad)^{3\nu-4} \left[ \guncs{\dfrac{\gammad}{\gamma + \gammad}}
- \guncs{\dfrac{\gammad}{\gammai + \gammad}} \right],  \\
\label{eq:regrelaxcritgp}
t^4 \calF(\gammac \le \gamma < \gammap ,t < \tc) &= \left(\dfrac{t}{\tini} \right)^4
\left(\dfrac{\aini}{a}\right)^3 \tini^4\,
\calNini\negthickspace\left\{\left[\gamma + \gammad\left(1 -
  \dfrac{\tini}{t} \right) \right] t \right\} \nonumber \\ & + c
(\gamma + \gammad)^{3\nu-4} \left[ \guncs{\dfrac{\gammad}{\gamma + \gammad}}
- \guncs{\dfrac{\gammad}{\gamma + \gammad} \dfrac{\tini}{t}} \right],\\
\label{eq:rerelaxcritgc}
t^4\calF(\gammax \le \gamma < \gammac,t < \tc) &= \left(\dfrac{t}{\tini}
\right)^4 \left(\dfrac{\aini}{a}\right)^3 \tini^4\,
\calNini\negthickspace\left\{\left[\gamma + \gammad\left(1 -
  \dfrac{\tini}{t} \right) \right] t \right\} \nonumber \\ & +
\dfrac{\cc}{\muc} (\gamma + \gammad)^{2 \chic - 3}
\left[\funcc{\dfrac{\gammad}{\gamma + \gammad}} -
  \left(\dfrac{\gamma+\gammad}{\gammac+\gammad}\right)^{\muc}
  \funcc{\dfrac{\gammad}{\gammac+\gammad}} \right]\nonumber \\ & +
c (\gamma + \gammad)^{3\nu-4} \left[\guncs{\dfrac{\gammad}{\gammac+\gammad}}  -
\guncs{\dfrac{\gammad}{\gamma + \gammad} \dfrac{\tini}{t}} \right],  \\
\label{eq:regrelaxcritgo}
t^4 \calF(0<\gamma < \gammax,t < \tc) &= \left(\dfrac{t}{\tini} \right)^4
\left(\dfrac{\aini}{a}\right)^3 \tini^4\,
\calNini\negthickspace\left\{\left[\gamma + \gammad\left(1 -
  \dfrac{\tini}{t} \right) \right] t \right\}\nonumber \\ & +
\dfrac{\cc}{\muc} (\gamma + \gammad)^{2 \chic - 3}
\funcc{\dfrac{\gammad}{\gamma + \gammad}} \nonumber \\ & -
\dfrac{\cc}{\muc} (\gamma + \gammad)^{2\chic - 3}
\left(\dfrac{t}{\tini} \right)^{-\muc} \funcc{\dfrac{\gammad}{\gamma +
    \gammad} \dfrac{\tini}{t}}.
\end{align}
\endgroup

Finally, for times $t \ge \tc$, $\gammap(t)$ becomes smaller than
$\gammac$ and the complete critical IR-regularised loop distribution
reads
\begingroup
\allowdisplaybreaks
\begin{align}
\label{eq:altregrelaxcritgi}
  t^4 \calF(\gamma \ge \gammai ,t \ge \tc ) &=  \left(\dfrac{t}{\tini} \right)^4
  \left(\dfrac{\aini}{a}\right)^3 \tini^4\,
  \calNini\negthickspace\left\{\left[\gamma + \gammad\left(1 -
    \dfrac{\tini}{t} \right) \right] t \right\}, \\
  \label{eq:altregcritgd}
t^4 \calF(\gammac \le \gamma < \gammai ,t \ge \tc ) &= \left(\dfrac{t}{\tini} \right)^4
\left(\dfrac{\aini}{a}\right)^3 \tini^4\,
\calNini\negthickspace\left\{\left[\gamma + \gammad\left(1 -
  \dfrac{\tini}{t} \right) \right] t \right\} \nonumber \\ & + c
(\gamma + \gammad)^{3\nu-4} \left[ \guncs{\dfrac{\gammad}{\gamma + \gammad}}
- \guncs{\dfrac{\gammad}{\gammai + \gammad}} \right],  \\
\label{eq:altregcritgc}
t^4 \calF(\gammap \le \gamma < \gammac ,t \ge \tc) &= \left(\dfrac{t}{\tini}
\right)^4 \left(\dfrac{\aini}{a}\right)^3 \tini^4\,
\calNini\negthickspace\left\{\left[\gamma + \gammad\left(1 -
  \dfrac{\tini}{t} \right) \right] t \right\} \nonumber \\ & +
\dfrac{\cc}{\muc} (\gamma + \gammad)^{2 \chic - 3}
\left[\funcc{\dfrac{\gammad}{\gamma + \gammad}} -
  \left(\dfrac{\gamma+\gammad}{\gammac+\gammad}\right)^{\muc}
  \funcc{\dfrac{\gammad}{\gammac+\gammad}} \right]\nonumber \\ & +
c (\gamma + \gammad)^{3\nu-4} \left[\guncs{\dfrac{\gammad}{\gammac+\gammad}}  -
\guncs{\dfrac{\gammad}{\gammai + \gammad}} \right],\\
\label{eq:altregcritgp}
t^4\calF(\gammax \le \gamma < \gammap,t \ge \tc) &= \left(\dfrac{t}{\tini}
\right)^4 \left(\dfrac{\aini}{a}\right)^3 \tini^4\,
\calNini\negthickspace\left\{\left[\gamma + \gammad\left(1 -
  \dfrac{\tini}{t} \right) \right] t \right\} \nonumber \\ & +
\dfrac{\cc}{\muc} (\gamma + \gammad)^{2 \chic - 3}
\left[\funcc{\dfrac{\gammad}{\gamma + \gammad}} -
  \left(\dfrac{\gamma+\gammad}{\gammac+\gammad}\right)^{\muc}
  \funcc{\dfrac{\gammad}{\gammac+\gammad}} \right]\nonumber \\ & +
c (\gamma + \gammad)^{3\nu-4} \left[\guncs{\dfrac{\gammad}{\gammac+\gammad}}  -
\guncs{\dfrac{\gammad}{\gamma + \gammad} \dfrac{\tini}{t}} \right],  \\
\label{eq:altregcritgo}
t^4 \calF(0<\gamma < \gammax,t \ge \tc) &= \left(\dfrac{t}{\tini} \right)^4
\left(\dfrac{\aini}{a}\right)^3 \tini^4\,
\calNini\negthickspace\left\{\left[\gamma + \gammad\left(1 -
  \dfrac{\tini}{t} \right) \right] t \right\}\nonumber \\ & +
\dfrac{\cc}{\muc} (\gamma + \gammad)^{2 \chic - 3}
\funcc{\dfrac{\gammad}{\gamma + \gammad}} \nonumber \\ & -
\dfrac{\cc}{\muc} (\gamma + \gammad)^{2\chic - 3}
\left(\dfrac{t}{\tini} \right)^{-\muc} \funcc{\dfrac{\gammad}{\gamma +
    \gammad} \dfrac{\tini}{t}}.
\end{align}
\endgroup

\bibliographystyle{JHEP}

\bibliography{strings}

\providecommand{\href}[2]{#2}\begingroup\raggedright\begin{thebibliography}{10}

\bibitem{Kirzhnits:1972}
D.~{Kirzhnits} and A.~{Linde}, \emph{{Macroscopic consequences of the Weinberg
  model}}, \href{http://dx.doi.org/10.1016/0370-2693(72)90109-8}{\emph{Phys.
  Lett. B} {\bf 42} (Dec., 1972) 471--474}.

\bibitem{Kibble:1976}
T.~W.~B. {Kibble}, \emph{{Topology of cosmic domains and strings.}}, {\emph{J.
  Phys. A} {\bf 9} (1976) 1387--1398}.

\bibitem{Witten:1985fp}
E.~Witten, \emph{{Cosmic Superstrings}},
  \href{http://dx.doi.org/10.1016/0370-2693(85)90540-4}{\emph{Phys. Lett.} {\bf
  B153} (1985) 243}.

\bibitem{Dvali:1998pa}
G.~R. Dvali and S.~H.~H. Tye, \emph{Brane inflation}, {\emph{Phys. Lett.} {\bf
  B450} (1999) 72--82}, [\href{http://arxiv.org/abs/hep-ph/9812483}{{\tt
  hep-ph/9812483}}].

\bibitem{Hindmarsh:1994re}
M.~B. Hindmarsh and T.~W.~B. Kibble, \emph{Cosmic strings}, {\emph{Rept. Prog.
  Phys.} {\bf 58} (1995) 477--562},
  [\href{http://arxiv.org/abs/hep-ph/9411342}{{\tt hep-ph/9411342}}].

\bibitem{Vilenkin:2000jqa}
A.~Vilenkin and E.~P.~S. Shellard, \emph{{Cosmic Strings and Other Topological
  Defects}}.
\newblock Cambridge University Press, 2000.

\bibitem{Durrer:2002}
R.~{Durrer}, M.~{Kunz} and A.~{Melchiorri}, \emph{{Cosmic structure formation
  with topological defects}}, {\emph{Phys. Rep.} {\bf 364} (June, 2002) 1--81},
  [\href{http://arxiv.org/abs/astro-ph/0110348}{{\tt astro-ph/0110348}}].

\bibitem{Polchinski:2004ia}
J.~Polchinski, \emph{{Introduction to cosmic F- and D-strings}},
  \href{http://arxiv.org/abs/hep-th/0412244}{{\tt hep-th/0412244}}.

\bibitem{Davis:2008dj}
A.-C. Davis, P.~Brax and C.~van~de Bruck, \emph{{Brane Inflation and Defect
  Formation}}, \href{http://dx.doi.org/10.1098/rsta.2008.0065}{\emph{Phil.
  Trans. Roy. Soc. Lond.} {\bf A366} (2008) 2833--2842},
  [\href{http://arxiv.org/abs/0803.0424}{{\tt 0803.0424}}].

\bibitem{Copeland:2009ga}
E.~J. Copeland and T.~W.~B. Kibble, \emph{{Cosmic Strings and Superstrings}},
  \href{http://dx.doi.org/10.1098/rspa.2009.0591}{\emph{Proc. Roy. Soc. Lond.}
  {\bf A466} (2010) 623--657}, [\href{http://arxiv.org/abs/0911.1345}{{\tt
  0911.1345}}].

\bibitem{Sakellariadou:2009ev}
M.~Sakellariadou, \emph{{Cosmic Strings and Cosmic Superstrings}},
  \href{http://dx.doi.org/10.1016/j.nuclphysbps.2009.07.046}{\emph{Nucl. Phys.
  Proc. Suppl.} {\bf 192-193} (2009) 68--90},
  [\href{http://arxiv.org/abs/0902.0569}{{\tt 0902.0569}}].

\bibitem{Ringeval:2010ca}
C.~Ringeval, \emph{{Cosmic strings and their induced non-Gaussianities in the
  cosmic microwave background}}, {\emph{Adv. Astron.} {\bf 2010} (2010)
  380507}, [\href{http://arxiv.org/abs/1005.4842}{{\tt 1005.4842}}].

\bibitem{Vachaspati:2015cma}
T.~Vachaspati, L.~Pogosian and D.~Steer, \emph{{Cosmic Strings}},
  \href{http://dx.doi.org/10.4249/scholarpedia.31682}{\emph{Scholarpedia} {\bf
  10} (2015) 31682}, [\href{http://arxiv.org/abs/1506.04039}{{\tt
  1506.04039}}].

\bibitem{Ade:2013xla}
{\scshape Planck} collaboration, P.~A.~R. Ade et~al., \emph{{Planck 2013
  results. XXV. Searches for cosmic strings and other topological defects}},
  \href{http://dx.doi.org/10.1051/0004-6361/201321621}{\emph{Astron.
  Astrophys.} {\bf 571} (2014) A25},
  [\href{http://arxiv.org/abs/1303.5085}{{\tt 1303.5085}}].

\bibitem{Lazanu:2014eya}
A.~Lazanu and P.~Shellard, \emph{{Constraints on the Nambu-Goto cosmic string
  contribution to the CMB power spectrum in light of new temperature and
  polarisation data}},
  \href{http://dx.doi.org/10.1088/1475-7516/2015/02/024}{\emph{JCAP} {\bf 1502}
  (2015) 024}, [\href{http://arxiv.org/abs/1410.5046}{{\tt 1410.5046}}].

\bibitem{Lizarraga:2016onn}
J.~Lizarraga, J.~Urrestilla, D.~Daverio, M.~Hindmarsh and M.~Kunz, \emph{{New
  CMB constraints for Abelian Higgs cosmic strings}},
  \href{http://dx.doi.org/10.1088/1475-7516/2016/10/042}{\emph{JCAP} {\bf 1610}
  (2016) 042}, [\href{http://arxiv.org/abs/1609.03386}{{\tt 1609.03386}}].

\bibitem{2017MNRAS.472.4081M}
J.~D. {McEwen}, S.~M. {Feeney}, H.~V. {Peiris}, Y.~{Wiaux}, C.~{Ringeval} and
  F.~R. {Bouchet}, \emph{{Wavelet-Bayesian inference of cosmic strings embedded
  in the cosmic microwave background}},
  \href{http://dx.doi.org/10.1093/mnras/stx2268}{\emph{\mnras} {\bf 472} (Dec.,
  2017) 4081--4098}, [\href{http://arxiv.org/abs/1611.10347}{{\tt
  1611.10347}}].

\bibitem{Sadr:2017hfm}
A.~Vafaei~Sadr, S.~M.~S. Movahed, M.~Farhang, C.~Ringeval and F.~R. Bouchet,
  \emph{{Multi-Scale Pipeline for the Search of String-Induced CMB
  Anisotropies}}, \href{http://dx.doi.org/10.1093/mnras/stx3126}{\emph{Mon.
  Not. Roy. Astron. Soc.} {\bf 475} (2018) 1010--1022},
  [\href{http://arxiv.org/abs/1710.00173}{{\tt 1710.00173}}].

\bibitem{2019MNRAS.483.5179C}
R.~{Ciuca} and O.~F. {Hern{\'a}ndez}, \emph{{Inferring cosmic string tension
  through the neural network prediction of string locations in CMB maps}},
  \href{http://dx.doi.org/10.1093/mnras/sty3478}{\emph{\mnras} {\bf 483} (Mar.,
  2019) 5179--5187}, [\href{http://arxiv.org/abs/1810.11889}{{\tt
  1810.11889}}].

\bibitem{Ringeval:2017eww}
C.~Ringeval and T.~Suyama, \emph{{Stochastic gravitational waves from cosmic
  string loops in scaling}},
  \href{http://dx.doi.org/10.1088/1475-7516/2017/12/027}{\emph{JCAP} {\bf 1712}
  (2017) 027}, [\href{http://arxiv.org/abs/1709.03845}{{\tt 1709.03845}}].

\bibitem{Blanco-Pillado:2017rnf}
J.~J. Blanco-Pillado, K.~D. Olum and X.~Siemens, \emph{{New limits on cosmic
  strings from gravitational wave observation}},
  \href{http://dx.doi.org/10.1016/j.physletb.2018.01.050}{\emph{Phys. Lett.}
  {\bf B778} (2018) 392--396}, [\href{http://arxiv.org/abs/1709.02434}{{\tt
  1709.02434}}].

\bibitem{Abbott:2017mem}
{\scshape LIGO Scientific, Virgo} collaboration, B.~Abbott et~al.,
  \emph{{Constraints on cosmic strings using data from the first Advanced LIGO
  observing run}},
  \href{http://dx.doi.org/10.1103/PhysRevD.97.102002}{\emph{Phys. Rev.} {\bf
  D97} (2018) 102002}, [\href{http://arxiv.org/abs/1712.01168}{{\tt
  1712.01168}}].

\bibitem{Ringeval:2005kr}
C.~Ringeval, M.~Sakellariadou and F.~Bouchet, \emph{{Cosmological evolution of
  cosmic string loops}}, {\emph{JCAP} {\bf 0702} (2007) 023},
  [\href{http://arxiv.org/abs/astro-ph/0511646}{{\tt astro-ph/0511646}}].

\bibitem{Vanchurin:2005pa}
V.~Vanchurin, K.~D. Olum and A.~Vilenkin, \emph{{Scaling of cosmic string
  loops}}, \href{http://dx.doi.org/10.1103/PhysRevD.74.063527}{\emph{Phys.
  Rev.} {\bf D74} (2006) 063527},
  [\href{http://arxiv.org/abs/gr-qc/0511159}{{\tt gr-qc/0511159}}].

\bibitem{Martins:2005es}
C.~J. A.~P. Martins and E.~P.~S. Shellard, \emph{{Fractal properties and
  small-scale structure of cosmic string networks}},
  \href{http://dx.doi.org/10.1103/PhysRevD.73.043515}{\emph{Phys. Rev.} {\bf
  D73} (2006) 043515}, [\href{http://arxiv.org/abs/astro-ph/0511792}{{\tt
  astro-ph/0511792}}].

\bibitem{Vincent:1998}
G.~{Vincent}, N.~D. {Antunes} and M.~{Hindmarsh}, \emph{{Numerical Simulations
  of String Networks in the Abelian-Higgs Model}}, {\emph{Phys. Rev. Lett.}
  {\bf 80} (Mar., 1998) 2277--2280},
  [\href{http://arxiv.org/abs/hep-ph/9708427}{{\tt hep-ph/9708427}}].

\bibitem{Moore:2002}
J.~N. {Moore}, E.~P.~S. {Shellard} and C.~J.~A.~P. {Martins}, \emph{{Evolution
  of Abelian-Higgs string networks}}, {\emph{Phys. Rev.} {\bf D65} (Jan., 2001)
  023503}, [\href{http://arxiv.org/abs/hep-ph/0107171}{{\tt hep-ph/0107171}}].

\bibitem{Hindmarsh:2008dw}
M.~Hindmarsh, S.~Stuckey and N.~Bevis, \emph{{Abelian Higgs Cosmic Strings:
  Small Scale Structure and Loops}},
  \href{http://dx.doi.org/10.1103/PhysRevD.79.123504}{\emph{Phys. Rev.} {\bf
  D79} (2009) 123504}, [\href{http://arxiv.org/abs/0812.1929}{{\tt
  0812.1929}}].

\bibitem{Hindmarsh:2017qff}
M.~Hindmarsh, J.~Lizarraga, J.~Urrestilla, D.~Daverio and M.~Kunz,
  \emph{{Scaling from gauge and scalar radiation in Abelian Higgs string
  networks}},  \href{http://arxiv.org/abs/1703.06696}{{\tt 1703.06696}}.

\bibitem{Blanco-Pillado:2013qja}
J.~J. Blanco-Pillado, K.~D. Olum and B.~Shlaer, \emph{{The number of cosmic
  string loops}},
  \href{http://dx.doi.org/10.1103/PhysRevD.89.023512}{\emph{Phys.Rev.} {\bf
  D89} (2014) 023512}, [\href{http://arxiv.org/abs/1309.6637}{{\tt
  1309.6637}}].

\bibitem{Blanco-Pillado:2017oxo}
J.~J. Blanco-Pillado and K.~D. Olum, \emph{{Stochastic gravitational wave
  background from smoothed cosmic string loops}},
  \href{http://dx.doi.org/10.1103/PhysRevD.96.104046}{\emph{Phys. Rev.} {\bf
  D96} (2017) 104046}, [\href{http://arxiv.org/abs/1709.02693}{{\tt
  1709.02693}}].

\bibitem{Quashnock:1990wv}
J.~M. Quashnock and D.~N. Spergel, \emph{{Gravitational Selfinteractions of
  Cosmic Strings}},
  \href{http://dx.doi.org/10.1103/PhysRevD.42.2505}{\emph{Phys. Rev.} {\bf D42}
  (1990) 2505--2520}.

\bibitem{Helfer:2018qgv}
T.~Helfer, J.~C. Aurrekoetxea and E.~A. Lim, \emph{{Cosmic String Loop Collapse
  in Full General Relativity}},  \href{http://arxiv.org/abs/1808.06678}{{\tt
  1808.06678}}.

\bibitem{Vachaspati:1984gt}
T.~Vachaspati and A.~Vilenkin, \emph{{Gravitational Radiation from Cosmic
  Strings}}, \href{http://dx.doi.org/10.1103/PhysRevD.31.3052}{\emph{Phys.
  Rev.} {\bf D31} (1985) 3052}.

\bibitem{PhysRevD.45.1898}
B.~Allen and E.~P.~S. Shellard, \emph{Gravitational radiation from cosmic
  strings}, \href{http://dx.doi.org/10.1103/PhysRevD.45.1898}{\emph{Phys. Rev.
  D} {\bf 45} (Mar, 1992) 1898--1912}.

\bibitem{Wachter:2016rwc}
J.~M. Wachter and K.~D. Olum, \emph{{Gravitational backreaction on piecewise
  linear cosmic string loops}},
  \href{http://dx.doi.org/10.1103/PhysRevD.95.023519}{\emph{Phys. Rev.} {\bf
  D95} (2017) 023519}, [\href{http://arxiv.org/abs/1609.01685}{{\tt
  1609.01685}}].

\bibitem{Copeland:1998na}
E.~J. Copeland, T.~W.~B. Kibble and D.~A. Steer, \emph{{The evolution of a
  network of cosmic string loops}},
  \href{http://dx.doi.org/10.1103/PhysRevD.58.043508}{\emph{Phys. Rev.} {\bf
  D58} (1998) 043508}, [\href{http://arxiv.org/abs/hep-ph/9803414}{{\tt
  hep-ph/9803414}}].

\bibitem{Rocha:2007ni}
J.~V. Rocha, \emph{{Scaling solution for small cosmic string loops}},
  \href{http://dx.doi.org/10.1103/PhysRevLett.100.071601}{\emph{Phys. Rev.
  Lett.} {\bf 100} (2008) 071601}, [\href{http://arxiv.org/abs/0709.3284}{{\tt
  0709.3284}}].

\bibitem{Lorenz:2010sm}
L.~Lorenz, C.~Ringeval and M.~Sakellariadou, \emph{{Cosmic string loop
  distribution on all length scales and at any redshift}},
  \href{http://dx.doi.org/10.1088/1475-7516/2010/10/003}{\emph{JCAP} {\bf 1010}
  (2010) 003}, [\href{http://arxiv.org/abs/1006.0931}{{\tt 1006.0931}}].

\bibitem{Vanchurin:2011hm}
V.~Vanchurin, \emph{{Towards a kinetic theory of strings}},
  \href{http://dx.doi.org/10.1103/PhysRevD.83.103525}{\emph{Phys.Rev.} {\bf
  D83} (2011) 103525}, [\href{http://arxiv.org/abs/1103.1593}{{\tt
  1103.1593}}].

\bibitem{Peter:2013jj}
P.~Peter and C.~Ringeval, \emph{{A Boltzmann treatment for the vorton excess
  problem}}, \href{http://dx.doi.org/10.1088/1475-7516/2013/05/005}{\emph{JCAP}
  {\bf 1305} (2013) 005}, [\href{http://arxiv.org/abs/1302.0953}{{\tt
  1302.0953}}].

\bibitem{Vanchurin:2013tk}
V.~Vanchurin, \emph{{Kinetic Theory and Hydrodynamics of Cosmic Strings}},
  \href{http://dx.doi.org/10.1103/PhysRevD.87.063508,
  10.1103/PhysRevD.87.069910}{\emph{Phys. Rev.} {\bf D87} (2013) 063508},
  [\href{http://arxiv.org/abs/1301.1973}{{\tt 1301.1973}}].

\bibitem{Schubring:2013xza}
D.~Schubring and V.~Vanchurin, \emph{{Transport Equation for Nambu-Goto
  Strings}}, \href{http://dx.doi.org/10.1103/PhysRevD.89.083530}{\emph{Phys.
  Rev.} {\bf D89} (2014) 083530}, [\href{http://arxiv.org/abs/1310.6763}{{\tt
  1310.6763}}].

\bibitem{Caldwell:1991jj}
R.~R. Caldwell and B.~Allen, \emph{{Cosmological constraints on cosmic string
  gravitational radiation}},
  \href{http://dx.doi.org/10.1103/PhysRevD.45.3447}{\emph{Phys. Rev.} {\bf D45}
  (1992) 3447--3468}.

\bibitem{Damour:2000wa}
T.~Damour and A.~Vilenkin, \emph{{Gravitational wave bursts from cosmic
  strings}}, \href{http://dx.doi.org/10.1103/PhysRevLett.85.3761}{\emph{Phys.
  Rev. Lett.} {\bf 85} (2000) 3761--3764},
  [\href{http://arxiv.org/abs/gr-qc/0004075}{{\tt gr-qc/0004075}}].

\bibitem{DePies:2007bm}
M.~R. DePies and C.~J. Hogan, \emph{{Stochastic Gravitational Wave Background
  from Light Cosmic Strings}},
  \href{http://dx.doi.org/10.1103/PhysRevD.75.125006}{\emph{Phys. Rev.} {\bf
  D75} (2007) 125006}, [\href{http://arxiv.org/abs/astro-ph/0702335}{{\tt
  astro-ph/0702335}}].

\bibitem{Regimbau:2011bm}
T.~Regimbau, S.~Giampanis, X.~Siemens and V.~Mandic, \emph{{The stochastic
  background from cosmic (super)strings: popcorn and (Gaussian) continuous
  regimes}}, \href{http://dx.doi.org/10.1103/PhysRevD.85.066001,
  10.1103/PhysRevD.85.069902}{\emph{Phys. Rev.} {\bf D85} (2012) 066001},
  [\href{http://arxiv.org/abs/1111.6638}{{\tt 1111.6638}}].

\bibitem{Binetruy:2012ze}
P.~Binetruy, A.~Bohe, C.~Caprini and J.-F. Dufaux, \emph{{Cosmological
  Backgrounds of Gravitational Waves and eLISA/NGO: Phase Transitions, Cosmic
  Strings and Other Sources}},
  \href{http://dx.doi.org/10.1088/1475-7516/2012/06/027}{\emph{JCAP} {\bf 1206}
  (2012) 027}, [\href{http://arxiv.org/abs/1201.0983}{{\tt 1201.0983}}].

\bibitem{Kuroyanagi:2012wm}
S.~Kuroyanagi, K.~Miyamoto, T.~Sekiguchi, K.~Takahashi and J.~Silk,
  \emph{{Forecast constraints on cosmic string parameters from gravitational
  wave direct detection experiments}},
  \href{http://dx.doi.org/10.1103/PhysRevD.86.023503}{\emph{Phys. Rev.} {\bf
  D86} (2012) 023503}, [\href{http://arxiv.org/abs/1202.3032}{{\tt
  1202.3032}}].

\bibitem{Aasi:2013vna}
{\scshape VIRGO, LIGO Scientific} collaboration, J.~Aasi et~al.,
  \emph{{Constraints on cosmic strings from the LIGO-Virgo gravitational-wave
  detectors}},
  \href{http://dx.doi.org/10.1103/PhysRevLett.112.131101}{\emph{Phys. Rev.
  Lett.} {\bf 112} (2014) 131101}, [\href{http://arxiv.org/abs/1310.2384}{{\tt
  1310.2384}}].

\bibitem{Henrot-Versille:2014jua}
S.~Henrot-Versille et~al., \emph{{Improved constraint on the primordial
  gravitational-wave density using recent cosmological data and its impact on
  cosmic string models}},
  \href{http://dx.doi.org/10.1088/0264-9381/32/4/045003}{\emph{Class. Quant.
  Grav.} {\bf 32} (2015) 045003}, [\href{http://arxiv.org/abs/1408.5299}{{\tt
  1408.5299}}].

\bibitem{Sousa:2016ggw}
L.~Sousa and P.~P. Avelino, \emph{{Probing Cosmic Superstrings with
  Gravitational Waves}},
  \href{http://dx.doi.org/10.1103/PhysRevD.94.063529}{\emph{Phys. Rev.} {\bf
  D94} (2016) 063529}, [\href{http://arxiv.org/abs/1606.05585}{{\tt
  1606.05585}}].

\bibitem{Bennett:1989}
D.~P. {Bennett} and F.~R. {Bouchet}, \emph{{Cosmic-string evolution}},
  {\emph{Phys. Rev. Lett.} {\bf 63} (Dec., 1989) 2776--2779}.

\bibitem{Bennett:1990}
D.~P. {Bennett} and F.~R. {Bouchet}, \emph{{High-resolution simulations of
  cosmic-string evolution. I. Network evolution}}, {\emph{Phys. Rev.} {\bf D41}
  (Apr., 1990) 2408--2433}.

\bibitem{Allen:1990}
B.~{Allen} and P.~{Shellard}, \emph{{Cosmic-string evolution - A numerical
  simulation}}, {\emph{Phys. Rev. Lett.} {\bf 64} (Jan., 1990) 119--122}.

\bibitem{Albrecht:1989}
A.~{Albrecht} and N.~{Turok}, \emph{{Evolution of cosmic string networks}},
  {\emph{Phys. Rev.} {\bf D40} (Aug., 1989) 973--1001}.

\bibitem{Sakellariadou:1990nd}
M.~Sakellariadou and A.~Vilenkin, \emph{{Cosmic-string evolution in flat
  space-time}}, \href{http://dx.doi.org/10.1103/PhysRevD.42.349}{\emph{Phys.
  Rev.} {\bf D42} (1990) 349--353}.

\bibitem{Copeland:2009dk}
E.~J. Copeland and T.~W.~B. Kibble, \emph{{Kinks and small-scale structure on
  cosmic strings}},
  \href{http://dx.doi.org/10.1103/PhysRevD.80.123523}{\emph{Phys. Rev.} {\bf
  D80} (2009) 123523}, [\href{http://arxiv.org/abs/0909.1960}{{\tt
  0909.1960}}].

\bibitem{Austin:1993rg}
D.~Austin, E.~J. Copeland and T.~W.~B. Kibble, \emph{{Evolution of cosmic
  string configurations}},
  \href{http://dx.doi.org/10.1103/PhysRevD.48.5594}{\emph{Phys. Rev.} {\bf D48}
  (1993) 5594--5627}, [\href{http://arxiv.org/abs/hep-ph/9307325}{{\tt
  hep-ph/9307325}}].

\bibitem{Polchinski:2006ee}
J.~Polchinski and J.~V. Rocha, \emph{Analytic study of small scale structure on
  cosmic strings}, {\emph{Phys. Rev.} {\bf D74} (2006) 083504},
  [\href{http://arxiv.org/abs/hep-ph/0606205}{{\tt hep-ph/0606205}}].

\bibitem{Dubath:2007mf}
F.~Dubath, J.~Polchinski and J.~V. Rocha, \emph{{Cosmic String Loops, Large and
  Small}}, \href{http://dx.doi.org/10.1103/PhysRevD.77.123528}{\emph{Phys.
  Rev.} {\bf D77} (2008) 123528}, [\href{http://arxiv.org/abs/0711.0994}{{\tt
  0711.0994}}].

\bibitem{Hindmarsh:2009qk}
M.~Hindmarsh, C.~Ringeval and T.~Suyama, \emph{{The CMB temperature bispectrum
  induced by cosmic strings}},
  \href{http://dx.doi.org/10.1103/PhysRevD.80.083501}{\emph{Phys. Rev.} {\bf
  D80} (2009) 083501}, [\href{http://arxiv.org/abs/0908.0432}{{\tt
  0908.0432}}].

\bibitem{Hindmarsh:2009es}
M.~Hindmarsh, C.~Ringeval and T.~Suyama, \emph{{The CMB temperature trispectrum
  of cosmic strings}},
  \href{http://dx.doi.org/10.1103/PhysRevD.81.063505}{\emph{Phys. Rev.} {\bf
  D81} (2010) 063505}, [\href{http://arxiv.org/abs/0911.1241}{{\tt
  0911.1241}}].

\bibitem{Vachaspati:1984dz}
T.~Vachaspati and A.~Vilenkin, \emph{{Formation and Evolution of Cosmic
  Strings}}, \href{http://dx.doi.org/10.1103/PhysRevD.30.2036}{\emph{Phys.
  Rev.} {\bf D30} (1984) 2036}.

\bibitem{Polchinski:2007rg}
J.~Polchinski and J.~V. Rocha, \emph{{Cosmic string structure at the
  gravitational radiation scale}},
  \href{http://dx.doi.org/10.1103/PhysRevD.75.123503}{\emph{Phys. Rev.} {\bf
  D75} (2007) 123503}, [\href{http://arxiv.org/abs/gr-qc/0702055}{{\tt
  gr-qc/0702055}}].

\bibitem{Gradshteyn:1965aa}
I.~S. Gradshteyn and I.~M. Ryzhik, \emph{Table of Integrals, Series, and
  Products}.
\newblock Academic Press, New York and London, 1965.

\end{thebibliography}\endgroup

\end{document}